\documentclass[iop,apj,appendixfloats]{emulateapj}
\usepackage{apjfonts}
\usepackage{graphicx}
\usepackage{amsmath}
\usepackage{amssymb}
\usepackage{color}

\gdef\xx[#1]{\textcolor{red}{#1}}

\gdef\kms{km\,s$^{-1}$}
\gdef\msun{M$_{\odot}$}

\gdef\gal{Dragonfly~44}
\gdef\ma{mag\,arcsec$^{-2}$}

\newcommand{\GG}[1]{}

\lefthead{van Dokkum et al.}
\righthead{}
\slugcomment{}
\slugcomment{The Astrophysical Journal, in press}
\begin{document}

\newcommand\XXX[1]{{\textcolor{red}{\textbf{x\ #1\ x}}}}


\title{Spatially-resolved stellar kinematics
of the ultra diffuse galaxy Dragonfly 44\\
I. Observations, Kinematics,
and Cold Dark Matter Halo Fits}

\author{Pieter van Dokkum\altaffilmark{1},
Asher Wasserman\altaffilmark{2},
Shany Danieli\altaffilmark{1},
Roberto Abraham\altaffilmark{3},
Jean Brodie\altaffilmark{2},
Charlie Conroy\altaffilmark{4},
Duncan A.\ Forbes\altaffilmark{5},
Christopher Martin\altaffilmark{6},
Matt Matuszewski\altaffilmark{6},
Aaron J.\ Romanowsky\altaffilmark{2,7},
Alexa Villaume\altaffilmark{2}
\vspace{8pt}}

\altaffiltext{1}
{Astronomy Department, Yale University, 52 Hillhouse Ave,
New Haven, CT 06511, USA}
\altaffiltext{2}
{University of California Observatories, 1156 High Street, Santa
Cruz, CA 95064, USA}
\altaffiltext{3}
{Department of Astronomy \& Astrophysics, University of Toronto, 50 St.\ George
Street, Toronto, ON M5S 3H4, Canada}
\altaffiltext{4}
{Harvard-Smithsonian Center for Astrophysics, 60 Garden Street,
Cambridge, MA, USA}
\altaffiltext{5}
{Centre for Astrophysics and Supercomputing, Swinburne University, Hawthorn,
VIC 3122, Australia}
\altaffiltext{6}
{Cahill Center for Astrophysics, California Institute of Technology, 1216 East
California Boulevard, Mail code 278-17, Pasadena, CA 91125, USA}
\altaffiltext{7}
{Department of Physics and Astronomy, San Jos\'e State University,
San Jose, CA 95192, USA}

\begin{abstract}

We present spatially-resolved stellar kinematics of the well-studied
ultra diffuse galaxy (UDG) \gal,
as determined from 25.3\,hrs of observations
with the Keck Cosmic Web Imager.
The luminosity-weighted dispersion within the half-light radius is
$\sigma_{1/2} = 33^{+3}_{-3}$\,\kms, lower than what we had
inferred before from a DEIMOS spectrum in the H$\alpha$ region.
There is no evidence for rotation, with $V_{\rm max}/\langle{}\sigma\rangle<0.12$
(90\,\% confidence) along the major axis, in possible conflict with
models where
UDGs are the high-spin tail of the normal dwarf galaxy distribution.
The spatially-averaged line profile
is more peaked than a Gaussian,
with Gauss-Hermite coefficient $h_4 = 0.13 \pm 0.05$.
The mass-to-light ratio within the effective radius is
$(M_{\rm dyn}/L_I)(<R_{\rm e})=26^{+7}_{-6}$\,M$_{\odot}/$L$_{\odot}$, similar
to other UDGs and higher by a factor of six than
smaller galaxies of the same luminosity. 
This difference between UDGs and other galaxies
is, however, sensitive to the aperture that is used,
and is much reduced when the $M/L$ ratios are measured within a fixed
radius of 10 kpc. \gal\ has a rising velocity dispersion profile,
from $\sigma=26^{+4}_{-4}$\,\kms\ at $R=0.2$\,kpc
to $\sigma=41^{+8}_{-8}$\,\kms\ at $R=5.1$\,kpc.
The profile can only be fit with a cuspy NFW profile if the orbital
distribution has strong tangential anisotropy, with $\beta=-0.8^{+0.4}_{-0.5}$.
An alternative explanation is that the dark matter profile has a core:
a Di Cintio et al.\ (2014) density profile with a mass-dependent core provides
a very good fit to the kinematics for a halo mass
of $\log (M_{200}/{\rm M}_{\odot})=11.2^{+0.6}_{-0.6}$ and
$\beta = -0.1^{+0.2}_{-0.3}$, i.e.\ isotropic orbits.
This model predicts a slight positive kurtosis,
in qualitative agreement with the measured $h_4$ parameter.
UDGs such as \gal\ are dark matter dominated even in their centers,
and can constrain
the properties of dark matter in a regime where baryons usually dominate the kinematics: small spatial scales in massive halos.
In a companion paper (Wasserman et al.\ 2019) we provide constraints
on the axion mass in the context of ``fuzzy'' dark matter models.

\end{abstract}


\section{Introduction}

Over the past several years it has been found that large, quiescent
galaxies with very
low central surface brightness  are surprisingly common
({van Dokkum} {et~al.} 2015; {Koda} {et~al.} 2015; {van der Burg}, {Muzzin}, \&  {Hoekstra} 2016).
Ultra diffuse galaxies (UDGs), with half-light radii $R_{\rm e}\gtrsim 1.5$\,kpc
and central surface brightness $\mu(g,0)\gtrsim 24$\,\ma, dominate the population of
large galaxies in rich clusters ({Danieli} \& {van Dokkum} 2019) and have also been
found in groups and the general field ({Merritt} {et~al.} 2016; {Mart{\'{\i}}nez-Delgado} {et~al.} 2016; {Rom{\'a}n} \& {Trujillo} 2017; {van der Burg} {et~al.} 2017).

UDGs exhibit a wide variety of properties,
as might perhaps be expected given their broad selection criteria:
many are smooth and round, resembling very large dwarf spheroidals
({van Dokkum} {et~al.} 2015),
some are clearly tidally-disrupted (such as the
spectacular boomerang-shaped galaxy M101-DF4; {Merritt} {et~al.} 2016), and others are gas rich
with widely distributed low level star formation (e.g., {Leisman} {et~al.} 2017).
An intriguing aspects of UDGs is that they often have many globular
clusters. The number of clusters varies strongly from galaxy to galaxy,
but
on average it is 5--7 times higher than in other
galaxies of the same luminosity
({Beasley} {et~al.} 2016; {Peng} \& {Lim} 2016; {van Dokkum} {et~al.} 2017a; {Amorisco} {et~al.} 2018; {Lim} {et~al.} 2018; {Forbes} {et~al.} 2018).
In at least some UDGs
the clusters have similar colors to the smooth galaxy light, and 
in those UDGs both the clusters and the diffuse light appear to be
old, metal poor, and $\alpha-$enhanced, similar to
many globular
clusters in the Milky Way (e.g., {Beasley} \& {Trujillo} 2016; {Gu} {et~al.} 2018; {van Dokkum} {et~al.} 2018c).
Other UDGs appear to be younger, and
may have more complex histories
({Ferr{\'e}-Mateu} {et~al.} 2018; {Fensch} {et~al.} 2019; {Mart{\'{\i}}n-Navarro} {et~al.} 2019).

From a galaxy formation perspective UDGs pose an interesting challenge,
as their existence was not explicitly
predicted. There are ways to puff up galaxies after their initial formation,
for example through external tides ({Hayashi} {et~al.} 2003; {Yozin} \& {Bekki} 2015; {Ogiya} 2018; {Jiang} {et~al.} 2019; {Carleton} {et~al.} 2019; Martin et al.\ 2019) or strong
supernova feedback ({Agertz} \& {Kravtsov} 2016; {Di Cintio} {et~al.} 2017; {Chan} {et~al.} 2018).
Such ``processing'' scenarios likely play an important role,
although they do not easily account for the high globular cluster
numbers ({Lim} {et~al.} 2018) or the apparent structural integrity ({Mowla} {et~al.} 2017) of many UDGs
in the Coma cluster. Other models seek the origin of UDGs in a combination of
low mass,
high spin, and late formation ({Amorisco} \& {Loeb} 2016; {Rong} {et~al.} 2017,
{Liao} {et~al.} 2019). These models have
the benefit of explaining their ubiquity but generically
predict that UDGs are young disks, requiring additional processing
to turn them into old spheroidal objects (see Liao et al.\ 2019).
In {van Dokkum} {et~al.} (2018c) we suggested that globular cluster-rich UDGs 
had extremely high gas densities at the time of their formation, and that feedback
from an intense, compact star burst that created the globular clusters caused
both the cessation of star formation and the expansion of the
galaxies. However, this idea is little
more than speculation at this point (see {Katz} \& {Ricotti} 2013, for related ideas).

Somewhat irrespective of their structural evolution
and star formation history, UDGs may provide constraints on the
nature and spatial distribution of dark matter. As has long been recognized for
dwarf spheroidals ({Lin} \& {Faber} 1983; {Walker} {et~al.} 2007; {Battaglia} {et~al.} 2008; {Walker} \& {Pe{\~n}arrubia} 2011) and
low surface brightness gas-rich dwarfs and
spirals ({de Blok} {et~al.} 2001; {Swaters} {et~al.} 2003; {Hayashi} {et~al.} 2004),
galaxies with a low baryon density offer relatively unambiguous information
on the dark matter profile. These profiles are often found
to be shallower than the cuspy NFW form ({Navarro}, {Frenk}, \& {White} 1997). The origin
of these shallow profiles (cores) is not well understood; proposed
explanations include tidal effects
({Read} \& {Gilmore} 2005), baryonic processes
such as supernova feedback ({Governato} {et~al.} 2010; {Pontzen} \& {Governato} 2012; {Di Cintio} {et~al.} 2014b),
warm or mixed dark matter (see {Macci{\`o}} {et~al.} 2012), and ``fuzzy'' dark
matter (e.g., {Marsh} \& {Silk} 2014).
The latter idea postulates that the dark matter is an
ultra-light axion with a de Broglie wavelength of hundreds of parsecs.

In this context UDGs such as VCC\,1287 ({Beasley} {et~al.} 2016),
Dragonfly~17 ({Peng} \& {Lim} 2016), and Dragonfly~44 ({van Dokkum} {et~al.} 2016) 
occupy an interesting
region of parameter space, as their 
stellar masses  are a factor of $\sim 100$ higher than those
of dwarf spheroidal galaxies. If they
lie on or above the canonical halo mass -- stellar mass
relation  ({Moster}, {Naab}, \& {White} 2013) their halos masses are also much higher
than those of dwarf spheroidals, and they
could help disentangle the processes shaping the distribution of
dark matter on kpc scales.
An example is the formation of cores, as
tidal and baryonic explanations for their presence are
most effective at particular mass scales, possibly
around $M_{\rm halo}\sim 10^{11}$\,\msun\ ({Di Cintio} {et~al.} 2014b).
Similarly, in fuzzy dark matter
models the size of the central ``soliton'' is expected to scale with halo mass as
$r_{\rm sol} \propto M_{\rm halo}^{-1/3}$ (Schive et al.\ 2014; Wasserman et al.\ 2019),
which means that it
is expected to be a more distinct feature in the kinematic profiles of
more massive halos (see, e.g., {Hui} {et~al.} 2017). 

So far, only galaxy-integrated measurements of UDG kinematics have been
made, and only for a handful of galaxies. They paint a confusing
picture, and suggest a remarkable range of dark matter properties
(see {Spekkens} \& {Karunakaran} 2018; {Toloba} {et~al.} 2018; {Alabi} {et~al.} 2018; {Ferr{\'e}-Mateu} {et~al.} 2018).
At one extreme is the large
Coma UDG \gal, with a
stellar velocity dispersion of $\sigma=47^{+8}_{-6}$\,\kms,\footnote{At least according to {van Dokkum} {et~al.} (2016) -- the true value is almost certainly lower, as shown in this paper.}
$74\pm 18$ globular clusters, and an estimated halo mass of $M_{200}=10^{11}-
10^{12}$\,\msun\ ({van Dokkum} {et~al.} 2016; {Di Cintio} {et~al.} 2017). At the other are
the galaxies NGC\,1052-DF2 and NGC\,1052-DF4, with dispersions
of $\sigma=8.5^{+2.2}_{-3.1}$\,\kms\ ({Danieli} {et~al.} 2019) and
$\sigma=4.2^{+4.4}_{-2.2}$\,\kms\ ({van Dokkum} {et~al.} 2019) respectively.
These velocity dispersions are consistent with those expected from the stellar mass alone
(van Dokkum et al.\ 2018a; Wasserman et al.\ 2018a).
This large apparent difference in dark matter content is surprising as
\gal\ and the NGC\,1052 galaxies have very similar stellar masses and morphologies, and
are all rich in globular clusters. 

Here we present constraints on the mass {\em profile}
of a UDG, as derived from  spatially-resolved
kinematics. This measurement has recently become possible thanks to the
arrival of the Keck Cosmic Web Imager (KCWI; {Morrissey} {et~al.} 2012, 2018)
on the Keck\,II telescope. KCWI is a low surface brightness-optimized
integral field unit (IFU) spectrograph, and the combination of its
relatively high spectral resolution, low read noise, and blue wavelength
coverage make it a near-perfect instrument for UDG spectroscopy.
We chose the well-studied UDG \gal\ for our program (see {van Dokkum} {et~al.} 2017a).
This paper presents the observations, data analysis, kinematics, and
dark matter halo fits. Two companion papers discuss constraints on
the axion mass in fuzzy dark matter models (Wasserman et al.\ 2019) and
the stellar population of \gal\ (A.\ Villaume et al., in preparation).
We assume that \gal\ is at the distance of the Coma cluster,
and for convenience
we take 100\,Mpc for that distance (see {Carter} {et~al.} 2008).
All wavelengths are in air, not vacuum.

\begin{figure*}[htbp]
  \begin{center}
  \includegraphics[width=0.9\linewidth]{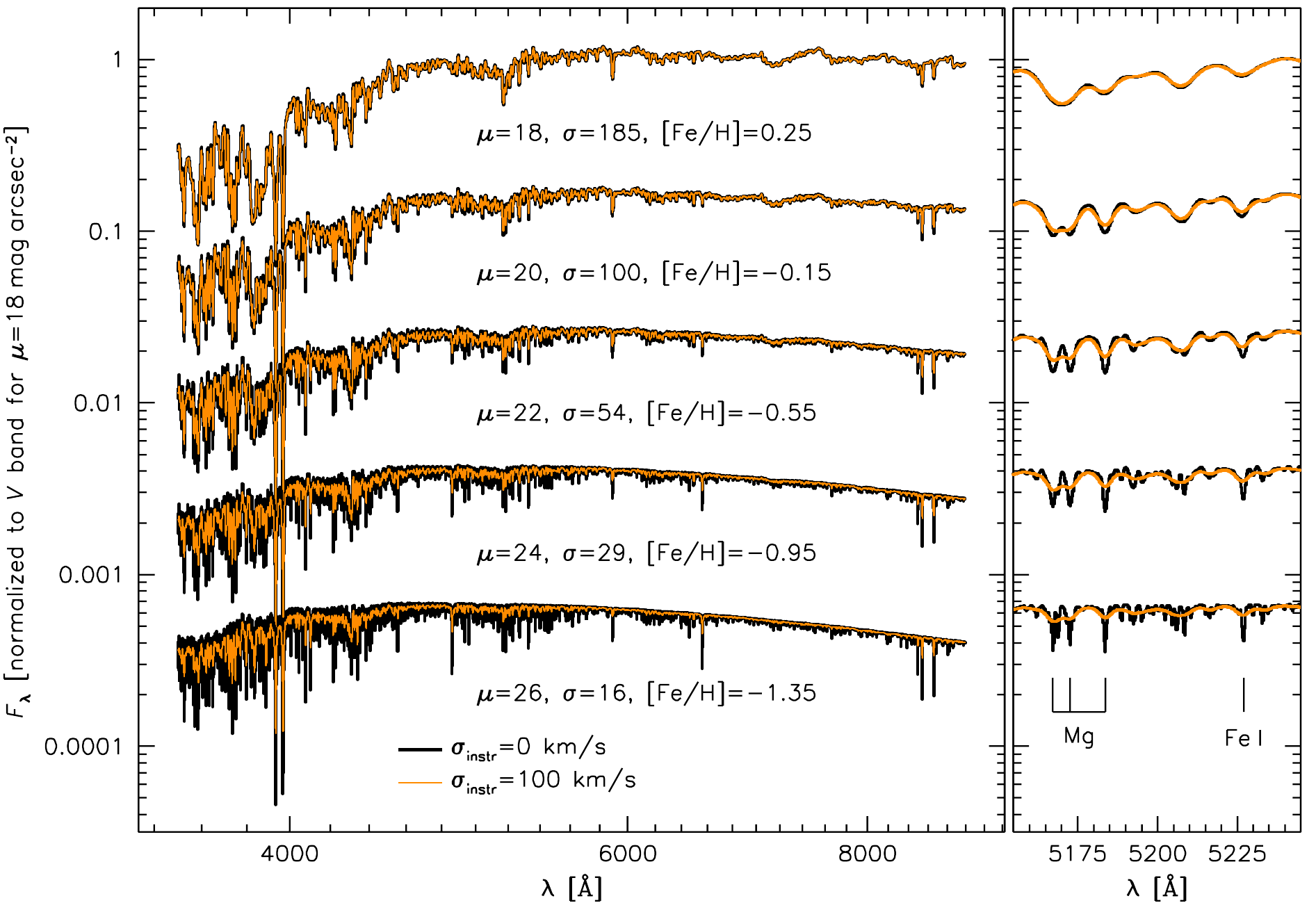}
  \end{center}
\vspace{-0.2cm}
    \caption{
Illustration of the expected optical
continuum spectra of old (age\,=\,10\,Gyr)
galaxies of decreasing central surface brightness,
from $\mu(0)\sim 18$\,\ma\ to $\mu(0)\sim 26$\,\ma.
The velocity dispersion (in \kms) and metallicity
of the model spectra are determined
from empirical relations (see text). The panel on the right shows the
region near 5200\,\AA. As metallicity and
velocity dispersion both decrease with decreasing surface brightness,
the observed depth of absorption features stays approximately constant as
long as the instrumental dispersion does not exceed the galaxy dispersion.
The orange spectra are for an instrumental resolution of
$\sigma_{\rm instr}=100$\,\kms, which is typical for low resolution / high
throughput spectrographs. At this resolution the low surface spectra
are nearly featureless, with the exception of the
intrinsically-broad Ca\,H+K lines in the near-UV.
}
\label{expect.fig}
\end{figure*}

\section{Observations}

\subsection{Expected integrated-light spectra for diffuse galaxies}
\label{expect_spec.sec}

We begin by discussing the {\em expected}
integrated-light spectra of quiescent low surface brightness galaxies,
as this is a relatively new topic
(see {van Dokkum} {et~al.} 2016; {Mart{\'{\i}}n-Navarro} {et~al.} 2019; {Emsellem} {et~al.} 2019; {Danieli} {et~al.} 2019; {Chilingarian} {et~al.} 2019).
Dynamical studies of low luminosity galaxies in the Local Group are typically
based on velocity measurements of individual stars.
As one example out of many,
{Geha} {et~al.} (2010) obtained the velocities of 520 stars in NGC\,147 and
442 stars in NGC\,185 to measure the velocity and velocity dispersion profiles
of these two Andromeda satellites out to $\sim 8$ effective radii.
They used the 
DEIMOS spectrograph on Keck, which offers excellent multiplexing
capability and a resolution of $\sigma_{\rm instr}
\approx 20$\,\kms\ in the Ca triplet region. Velocities are routinely determined
to an accuracy of 1--2\,\kms, enabling measurements of velocity dispersions down
to $\lesssim 5$\,\kms\ (e.g., {Ibata} {et~al.} 2006; {Martin} {et~al.} 2007; {Geha} {et~al.} 2010; {Collins} {et~al.} 2010). 

For low mass quiescent
galaxies beyond a few Mpc distance only integrated-light measurements
can be obtained, and velocity dispersions
can only be determined from the broadening of absorption lines
of the entire stellar population. This is difficult:
the surface brightness is low,
which means that the signal-to-noise (S/N) ratio per pixel is low; the metallicity is
low, which means that the metal lines are weak; and the velocity dispersion is low,
which means that the instrumental resolution needs to be relatively high.
The observed velocity dispersion
is related to the intrinsic velocity dispersion as
$\sigma_{\rm obs}^2 = \sigma_{\rm stars}^2 + \sigma_{\rm instr}^2$:
if the stellar dispersion is, say, 20\,\% of the instrumental resolution,
the observed dispersion is only 2\,\% larger than the instrumental broadening.
If $\sigma_{\rm stars}< \sigma_{\rm instr}$ systematic effects such as
template mismatch, small errors in the wavelength
calibration, and uncertainties in the (wavelength-dependent) spectral resolution
often dominate the error budget 
(see, e.g., {Kelson} {et~al.} 2000; Chilingarian et al.\ 2008; Strader et al.\ 2009;
Emsellem et al.\ 2019).\footnote{When
features are not well resolved, fitting codes effectively match the
total absorption (the product of the width and depth of the line) rather
than (just) the width. This is why template mismatch is a particularly
onerous problem when $\sigma_{\rm stars} < \sigma_{\rm instr}$: codes
appear to fit velocity widths, but what they are actually fitting is
equivalent widths.}

Fortunately, if the instrumental resolution is sufficiently high,
the low metallicity is compensated to some degree by the low intrinsic
velocity dispersion: they conspire to yield absorption features whose observed
strength is fairly independent of the surface brightness of the galaxy.
This is illustrated in Fig.\ \ref{expect.fig}, where we show the expected integrated-light
spectra of galaxies with a range of central surface brightness.
The spectra are synthetic stellar population synthesis models
({Conroy}, {Gunn}, \& {White} 2009), generated at
a resolution of $R=10,000$ ($\sigma_{\rm temp}= 13$\,\kms) using the
MIST isochrones ({Choi} {et~al.} 2016). We assume that the
velocity dispersion is related to the surface brightness as $\mu = 35 - 7.5\log \sigma$.
This relation is consistent with the central surface brightnesses and
velocity dispersions of elliptical galaxies (which have $\mu\sim 17.5$\,\ma\ and
$\sigma \sim 200$\,\kms; {Franx}, {Illingworth}, \&  {Heckman} 1989) and those of dwarf galaxies in the Local Group
($\mu\sim 27$\,\ma\ and $\sigma \sim 10$\,\kms; {McConnachie} 2012).
The relation between velocity dispersion and metallicity is obtained from an approximate
fit to the data in Fig.\ 12 of {Gu} {et~al.} (2018): [Fe/H]\,$= 1.5\log \sigma - 3.15$. The age
is assumed to be 10\,Gyr for all objects.

Black spectra are at the intrinsic resolution of the galaxies,\footnote{The
templates were smoothed by a Gaussian of width $\sigma_{\rm sm}^2=
\sigma_{\rm gal}^2-\sigma_{\rm temp}^2$.} that is, for a hypothetical
instrumental resolution of $\sigma_{\rm instr}=0$\,\kms. It is clear that even the
faintest galaxies with the lowest metallicity have many strong spectral features
in the optical, reaching a continuum absorption strength of $\sim 50$\,\%. However,
this is not the case at low spectral resolution. The orange lines show the same
spectra for $\sigma_{\rm instr}=100$\,\kms. At this instrumental resolution the features become
gradually weaker for fainter galaxies, with the Balmer lines and the Ca triplet
the only reasonably strong lines redward of $\lambda=4000$\,\AA\ for galaxies in the 
UDG regime ($\mu\gtrsim 24$\,\ma).

The models in Fig.\ \ref{expect.fig} demonstrate qualitatively
that it is possible to measure velocity dispersions from metal lines in
very low surface brightness galaxies, despite their low metallicity. The fact that
the lines are weak is compensated by the fact that they are not blended
and have maximum
absorption depths that are roughly independent of
surface brightness. Most studies of Local Group dwarfs measure stellar velocities
from H$\alpha$ and the Ca\,$\lambda\lambda 8662.1,8542.1,8498.0$
triplet lines, largely because the DEIMOS spectrograph
on Keck has its highest sensitivity and spectral resolution in the red.
However, at sufficiently high instrumental resolution
the integrated-light
spectrum blueward of $6000$\,\AA\ actually has the highest information content, as
is evident in Fig.\ \ref{expect.fig}.
The strongest features are the Ca\,$\lambda\lambda
3968.5,3933.7$ H+K lines. These intrinsically-broad lines cannot be used
for velocity dispersion measurements but are
probably the best features to target for redshift measurements of faint
low surface brightness objects.

\begin{figure*}[htbp]
  \begin{center}
  \includegraphics[width=1.0\linewidth]{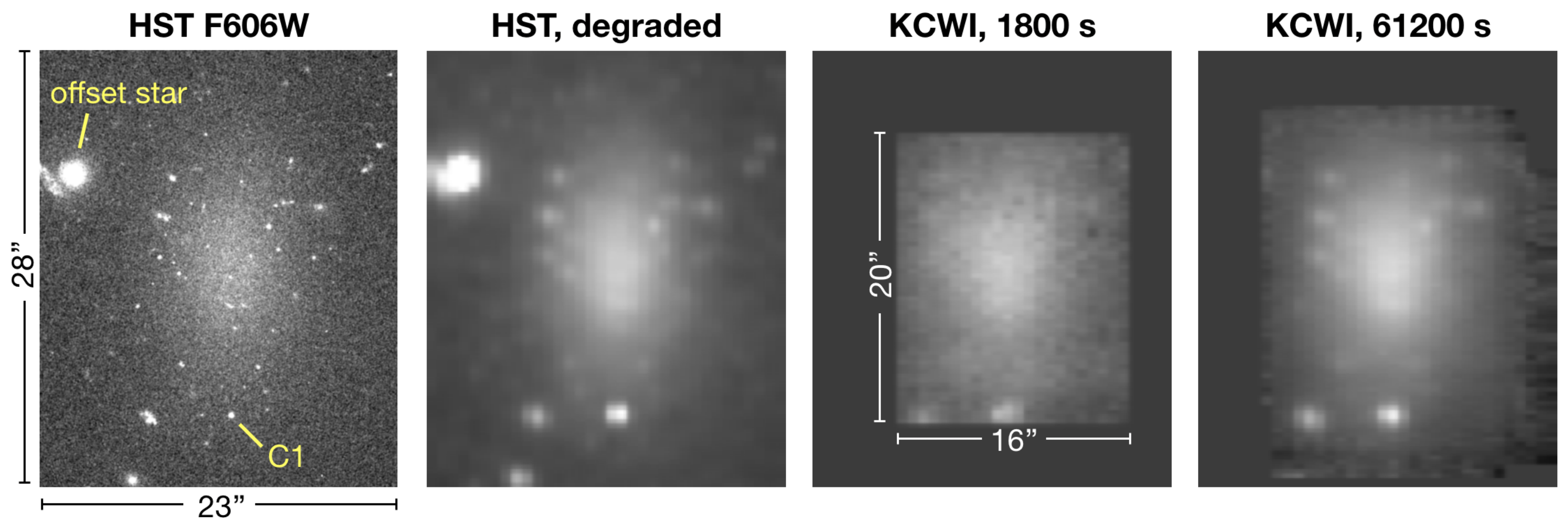}
  \end{center}
\vspace{-0.2cm}
    \caption{
{\em Left panel:} HST WFC3/UVIS
$V_{606}$ image of \gal, from {van Dokkum} {et~al.} (2017a), rotated by 32 degrees
with respect to North. The image spans 11.2\,kpc\,$\times$\,13.6\,kpc at the distance
of Coma. The offset star that was used to ensure accurate pointing is
marked, as well as a bright compact object (C1). {\em Second panel:} The HST image,
smoothed to ground-based seeing and sampled at the KCWI pixel scale. This model
for the KCWI data is used to align the individual KCWI exposures and to optimize the
background subtraction. {\em Third panel:} Individual KCWI science exposure (from
February 12 2018), after collapsing the data cube along the wavelength axis.
{\em Right panel:} Sum of aligned KCWI exposures, with a total exposure time of 17\,hrs.
}
\label{hst_kcwi.fig}
\end{figure*}

\subsection{Keck Cosmic Web Imager spectroscopy}
\label{obs.sec}

IFU spectroscopy of \gal\ was obtained with KCWI on Keck\,II in the
Spring of 2018, following initial observations in 2017 June
during commissioning of the instrument. The commissioning data informed the
observing strategy in 2018 but are not used in
the analysis:
conditions were variable, the observing strategy was not yet optimized, and
aspects of the instrument and data processing were still being finalized.
A list of 2018 dates and exposure times is provided in Table \ref{obs.table}.
The list does not include nights that had to be discarded due to cirrus or
clouds. Maunakea was plagued by bad weather during the entire
winter and spring of 2018, and the amount of useable time was about
one third of the total allocated time for this program.

\noindent
\begin{deluxetable}{lcc}
\tablecaption{Exposure times\label{obs.table}}
\tablehead{\colhead{Date} & \colhead{Science} & \colhead{Sky} \\
 & [sec] & [sec]}
\startdata
January 22 & 7,200 & 2,400 \\
February 11 & 10,800 & 4,800 \\
February 12 & 7,200 & 4,800 \\
February 13 & 10,800 & 4,800 \\
February 17 & 3,600 & 2,400 \\
February 18 & 5,400 & 2,400 \\
April 13 & 5,400 & 2,400 \\
April 17 & 1,800 & 2,400 \\
May 10 & 9,000 & 3,600 
\enddata
\end{deluxetable}
The medium slicer was used with the medium resolution BM grating,
for a field of view of $16\arcsec \times 20\arcsec$ and  an approximate
spectral resolution $R\sim 4000$ (see \S\,4.1 for a
measurement of the instrumental resolution as a function of wavelength).
The data were taken with $2\times 2$ binning to reduce read noise.
A sky position angle of $-32^{\arcdeg}$ was used, as this places the
major axis of the galaxy along the long ($20\arcsec$) axis of the
IFU. The field of view of KCWI is shown in
Fig.\ \ref{hst_kcwi.fig}, along with an HST WFC3/UVIS image
of \gal\ (from {van Dokkum} {et~al.} 2017a).
The central wavelength is $\lambda_{\rm cen}\approx 5050$\,\AA,
with small (10\,\AA\ -- 20\,\AA)
variations between observing nights so the same wavelengths
do not always fall on the same part of the detector. The effective
wavelength range is approximately $4650$\,\AA\ -- 5450\,\AA.

The observing strategy typically constituted of the following steps.
First a nearby star (see the left panel
of Fig.\ \ref{hst_kcwi.fig}) was acquired, using the
slit viewing camera to center the star in the KCWI field. Then a
pre-determined offset was applied to place \gal\ close
to the center of the field. This offset was varied by a few arcseconds
for each exposure; this is helpful for
diagnosing flat fielding and sky subtraction issues,
and yields data over a slightly larger area in the final combined
frames than the instantenous
KCWI field of view. We then obtained
a science exposure of 1,800\,s. After the first science exposure we
moved the telescope to a relatively empty area about $1\farcm 5$
away, and obtained a 1,200\,s ``sky'' exposure. These are crucial
for accurate sky subtraction, as \gal\ overfills the KCWI field
of view (see Fig.\ \ref{hst_kcwi.fig}). We then moved back to \gal\ and
obtained another 1,800\,s science exposure. A typical nightly
sequence was science -- sky -- science -- science -- sky -- science
-- science -- sky, but this varied somewhat during the runs due to changing
conditions, telescope/instrument problems, and other issues.

The total exposure time of frames that went into our final stack is
$61,200$\,s, or 17\,hrs. The total exposure
time that went into our sky analysis
is $30,000$\,s, and the total science + sky time that is used in the analysis
is 25.3 hrs. In addition to these science and blank field data we obtained
standard sets of daytime darks, flat fields, and arc lamp exposures.

\section{Data reduction}

\subsection{Pipeline processing}

The KCWI Data Extraction
and Reduction Pipeline (KDERP)
is maintained in a public {\tt github}
repository.\footnote{\url{https://github.com/Keck-DataReductionPipelines/KcwiDRP}}
We used this pipeline, with default settings, to turn individual science and sky frames
into wavelength-calibrated, flat-fielded, and cosmic ray-cleaned data cubes. What follows
is a brief summary of the pipeline processing steps; we refer to \S\,4
of {Morrissey} {et~al.} (2018) and the documentation
in the {\tt github} repository for more detailed information.

The pipeline is modular, with eight stages (nine when including a
``book keeping'' preparation step). In the first stage bias and overscan are
subtracted, the data are converted to electrons, and cosmic rays are removed using the
L.A.Cosmic ({van Dokkum} 2001) algorithm. The second stage subtracts the dark and
removes scattered light.
In the third stage analytic functions are found that describe
both geometric distortions and the
wavelength calibration. These are based on cross-correlations of the data
with  arc lamp
spectra and a pattern of continuum bars. The output of this stage
includes a map that provides the 3D data cube position for each pixel
in the 2D image. As discussed in \S\,\ref{solar.sec},
errors in the wavelength calibration
are $\approx 0.08$\,\AA, corresponding to $\approx 5$\,\kms. The fourth stage applies
a pixel-to-pixel flat field correction, as well as a correction
for vignetting and the overall illumination pattern.
Stage five is a
sky subtraction step, which in our analysis is carried out at a later
stage.
In stage six the data cubes are generated, based on
the functions that were derived in stage three. 
In stage seven the data are corrected for differential atmospheric refraction.
Stage eight is flux calibration, using a standard star; this stage is skipped in
our analysis as all our measurements are insensitive to the overall continuum
calibration.

The pipeline products that are used in the subsequent steps are the
``{\tt ocubed}'' files: the rectified, but un-skysubtracted cubes.
The data are sampled on a three-dimensional
grid with pixel size $0\farcs 68 \times 0\farcs 29 \times 0.5$\,\AA.

\subsection{Correction for residual spatial variation}

Inspection of the wavelength-collapsed 2D science and sky frames shows
a small gradient in the background at a level of $\approx 3$\,\%.
Within each night
the gradient has a similar amplitude (as a fraction of the background)
in the sky frames and the science frames, and
as these have different exposure times we conclude that it is likely
a multiplicative rather than an additive effect. For each night a
correction flat is created by fitting low order 2D
surfaces to all the wavelength-collapsed
sky frames and then averaging these fits. The fits are done iteratively
with aggressive outlier rejection so that serendipitous objects in
the sky frames are ignored. The science frames are then divided by the
correction flat. Although the amplitude varies somewhat from night
to night the
correction flats always show the same
pattern, a negative
gradient in the $x$-direction (the ``short'', $16\arcsec$, axis, which
corresponds to systematic slice-to-slice
variation over the whole detector rather than within slices).
We tested that the gradient is not driven by
a particular wavelength region, and
that treating the variation as an additive rather than a multiplicative
effect does not change the results.

\subsection{Alignment}

Spatially aligning the data cubes is not straightforward, as there is
no compact object that is bright enough to determine
an accurate position for every exposure. Object C1 (Fig.\ \ref{hst_kcwi.fig})
is usually near the edge of the field,
and in about half the exposures outside of it. 
Instead of using a single object we fit each collapsed data cube
to a 2D model of the flux in the entire KCWI field of view.
This model is created from the $V_{606}$ 
HST WFC3/UVIS image of \gal, shown in the left panel of Fig.\ \ref{hst_kcwi.fig}.
This image is convolved by a Gaussian with a FWHM of $1\farcs 0$ and
projected onto the same spatial grid (with a position angle
of $-32^{\arcdeg}$ and $0\farcs 68 \times 0\farcs 29$
pixels) as the KCWI data. The resulting model for the KCWI spatial
flux distribution is shown in the second panel of Fig.\ \ref{hst_kcwi.fig}.

Before performing the fit an approximate background is removed by
subtracting
the average of the flux in the outer 1-pixel wide perimeter of the
collapsed science frame.
Both the collapsed frame and the model are normalized so the total
flux in each
is 1.  The best fitting shift with respect to the model is found by a simple
grid search, subtracting the model from the shifted science frame at
each step and minimizing the square of the residuals.
All science exposures yield clear minima and stable solutions.
The data cubes are shifted to the common reference{} frame of the model using linear
interpolation. 

The final spatial resolution of the combined datacube is a reflection of the
seeing during the observations, guiding errors, and the uncertainties
in the alignment of individual exposures. We assess the spatial resolution
using object C1 in the summed, wavelength-collapsed data cube. After removing
the flux from the galaxy by fitting linear functions in $x$ and $y$, we
fit one-dimensional Gaussian profiles in both directions. The FWHM in
the $x$-direction is 1.98\,pixels, corresponding to $1\farcs 3$. The FWHM in the
better-sampled $y$ direction is 3.96 pixels, or $1\farcs 1$. At the distance
of Coma these values correspond to 0.63\,kpc and 0.53\,kpc respectively.

\subsection{Sky subtraction}

The sky subtraction is the most critical step in the data reduction.
The median galaxy signal ranges from $4.5$\,e$^{-}$\,pix$^{-1}$ 
in the center to 0.8\,e$^{-}$\,pix$^{-1}$
in the outer annulus, whereas the sky continuum is
$\approx 40$\,e$^{-}$\,pix$^{-1}$. Typical metal line absorption depths
of 10\,\% therefore correspond to 0.002\,$\times$ the sky brightness in the outer
annulus, which means that both emission and absorption features in the
sky spectrum need
to be modeled and subtracted with great precision.

\begin{figure*}[htbp]
  \begin{center}
  \includegraphics[width=0.95\linewidth]{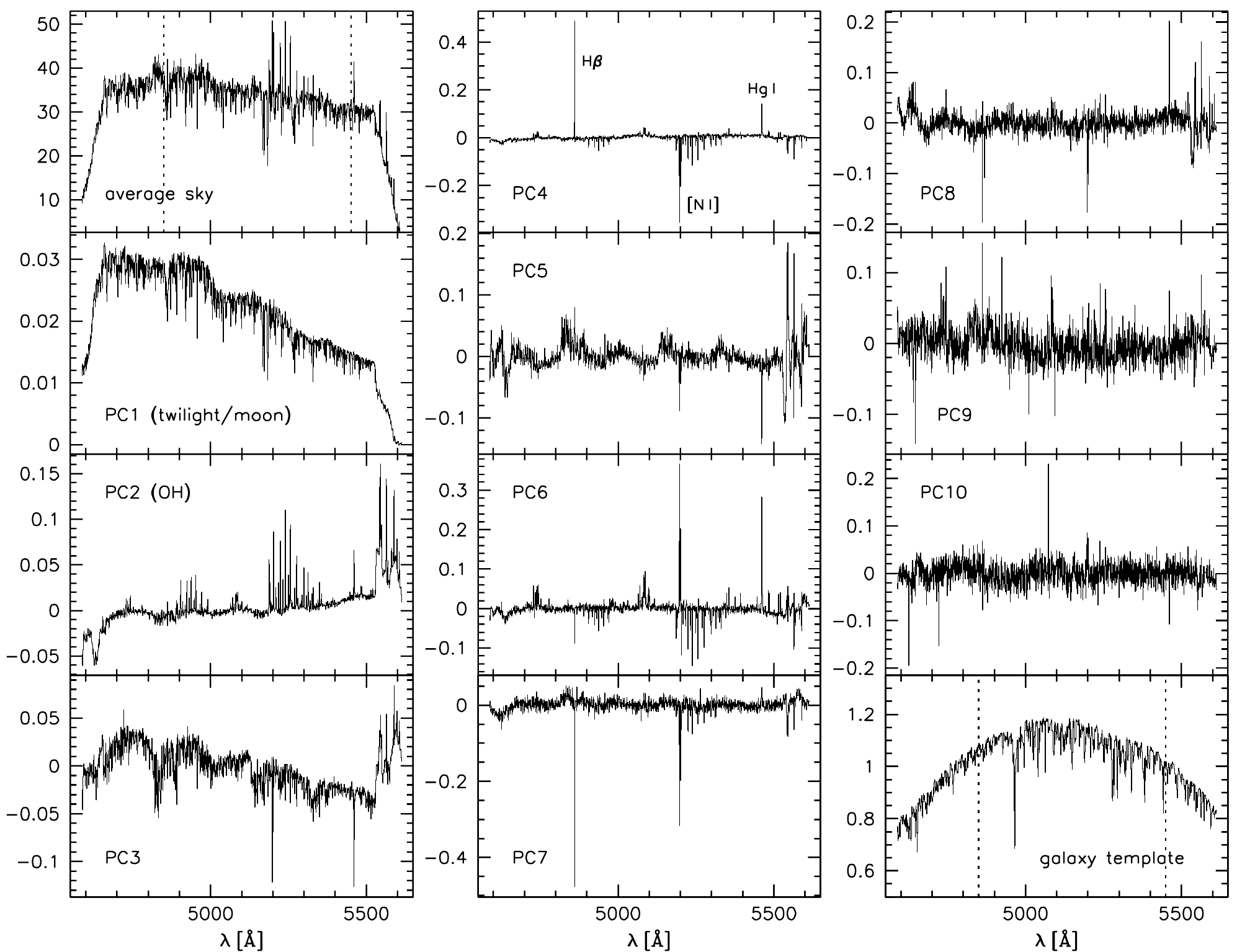}
  \end{center}
\vspace{-0.2cm}
    \caption{
Template spectra that are used to fit the background in the outer regions
of the science frames.
Along with the average blank sky spectrum (top left), ten PCA components
are shown along with a template for the ``contamination'' by light from
\gal\ (bottom right). The eigenspectra
PC1 -- PC10 describe the variation among the
individual sky spectra with
respect to the average. The key varying elements are the contribution
of the solar spectrum, OH bands, and several distinct emission lines
(particularly H$\beta$, [N\,{\sc i}], and Hg\,{\sc i}).
Dotted vertical lines indicate the spectral region that is used
for the kinematic measurements. Except for the (normalized)
galaxy template the units
of all spectra are e$^{-}$\,pix$^{-1}$.
}
\label{pca.fig}
\end{figure*}

\subsubsection{Principal component analysis}

A particular complication in our dataset is that \gal\ is larger than
the KCWI field of view, which means that we cannot determine the
sky spectrum from empty areas in the science cubes. We opted not to
use the nod-and-shuffle technique, as this comes with severe
penalties of a factor of two in S/N ratio (a factor of $\sqrt{2}$ due to
the fact that only 50\,\% of the exposure time is spent on-target,
and another factor of $\sqrt{2}$ because the noise from the offset
field is added to that of the science field) and a factor of four in
spectral coverage. Instead, as explained in \S\,\ref{obs.sec}
we interspersed the 1800\,s
science exposures with 1200\,s blank sky frames. We cannot
use an average of all the spatially-collapsed blank sky
spectra in a particular night to subtract the background
from the science cubes of that same night, as the sky
continuum, emission, and absorption all
vary too much within and between exposures. Instead, we use all the
spatially-collapsed individual sky spectra over multiple nights to capture the
wavelength-dependent {\em time variation} in the sky spectrum. This
variation is parameterized with a principal component analysis
(PCA), and the PCA components are then fitted to the background in each
of the science cubes. We note that this PCA analysis is different from
that used
in the
``Zurich Atmosphere Purge'' (ZAP; {Soto} {et~al.} 2016) algorithm; ZAP captures
any {\em spatial} variation in the sky spectrum caused by flat
fielding errors or other systematic effects, in regions away from
objects of interest.  In what follows
we describe our methodology in more detail.

We split the data in three overlapping
sets, as we find that the sky
variation in winter is somewhat different than in
spring. This could be a seasonal effect, but it is perhaps more
likely that it is due
to the fact that the time of night when
\gal\ is accessible changes during the Coma season
(the end of the night in January and the start of the night in May).
The first set consists of 14 sky exposures and the second and third consist
of 12 each. A 1D spectrum is created from each sky cube by collapsing
both spatial axes, after carefully masking all objects in the field.
Next, a PCA is performed on the 1D spectra within each set,
using the {\tt scikit-learn}
python implementation of singular value
decomposition.\footnote{\url{https://scikit-learn.org/}}
Ten components are used; we verified that
the results are nearly identical for eight, nine, or eleven.

The top left panel of Fig.\ \ref{pca.fig}
shows the average sky spectrum from the first set of
14 collapsed blank sky cubes.\footnote{Although they differ in the
details the other two sets produce very similar-looking
results.}  The spectrum is complex, with many absorption
and emission features reaching $\gtrsim 10$\,\% of the continuum.
The extremely strong [O\,{\sc i}]\,$\lambda 5577.3$ line was interpolated
over, as it falls outside of the wavelength range of interest (indicated
by the vertical dotted lines) and would otherwise dominate
many of the PCA components. The ten
eigenspectra are labeled PC1 -- PC10. They disentangle several distinct
causes of the variation in the night sky. PC1 is an excellent match to
the solar spectrum, as will be demonstrated in \S\,\ref{solar.sec}.
The contribution of the Sun is of course higher when data
are taken closer to morning
and evening twilight and also when the moon is above the horizon.

PC2 mostly reflects the variation in lines from OH radicals; these are
produced in the upper atmosphere from recombination of atomic oxygen
and are strongest near dawn and dusk.
The presence of these lines may seem surprising, as the OH
``forest'' is usually considered to only be present
at wavelengths $\lambda > 6100$\,\AA. At bluer
wavelengths the lines are weaker as they have small transition
probabilities, but as shown in Fig.\ \ref{pca.fig} the
8--1 and 9--2 Meinel bands
(see {Osterbrock} {et~al.} 1996, 2000) are important contributors to the sky
variation. Other important varying lines are H$\beta$
(from
geocoronal atomic hydrogen; {Burrage}, {Yee}, \& {Abreu} 1989), the
[N\,{\sc i}]\,$\lambda\lambda 5198.2,5200.5$ doublet ({Sharpee} {et~al.} 2005),
and Hg\,{\sc i}\,$\lambda$5460.7 ({Osterbrock} {et~al.} 2000).
These lines do not change in lockstep,
and the PC3 -- PC10 eigenspectra mostly capture different combinations of
positive and negative variations of the OH bands and the individual lines.

\subsubsection{Fitting and subtracting the sky in the science cubes}

The sky in each of the 34 science cubes is fitted with a linear
combination of templates. For each science cube, an
average ``sky + galaxy'' spectrum was created by averaging all pixels
after masking most of the light of \gal\ and
other objects in the field of view
(see \S\,\ref{apertures.sec}). Besides sky emission
this spectrum also contains flux from the galaxy, as \gal\ extends beyond
the KCWI field of view. In order to model this spectrum we maximize the
likelihood
\begin{equation}
\ln p = -\frac{1}{2}\sum_{\lambda=4700}^{5500}
\left[ \frac{\left(F_{\lambda} - M_{\lambda}
\right)^2}{e_{\lambda}^2} + \ln e_{\lambda}^2 \right],
\end{equation}
with $F$ the extracted sky + galaxy spectrum,
$e$ the errors in the data, and $M$ a model of the
form
\begin{equation}
M(\lambda) = \alpha_0 S_{\rm avg}(\lambda) + \sum_{i=1}^{10} \alpha_i {\rm PCA}_i(\lambda) 
 + \alpha_{11} T_{\rm gal}(\lambda).
\end{equation}
The model is a linear combination of the twelve templates shown in Fig.\
\ref{pca.fig}: the average sky spectrum, the ten eigenspectra describing
the variation in the sky, and a template for the contribution of \gal.
This galaxy template, $T_{\rm gal}$,
is created by redshifting a 10\,Gyr, [Fe/H]\,=\,$-1.0$
stellar population synthesis model (see \S\,\ref{expect_spec.sec}) to
the velocity of \gal,
convolving it to a resolution of 40\,\kms, and multiplying it by a low
order polynomial so that the continuum
roughly corresponds to that of the observed galaxy spectrum. This
last step accounts for the fact that
no flux calibration was done in the analysis.
These details do not influence the fit very much,
as the main function of $T_{\rm gal}$ is
to allow for an additive component that is not part of the sky spectrum.
As we show later \gal\ is actually not the only contribution of
this kind; there is an additional unidentified additive component in
the science data which is ``absorbed'' by $\alpha_{11}$, the coefficient
for the galaxy template.
In this context it is reassuring
that we find
nearly indistinguishable results for the best-fitting sky models
if $T_{\rm gal}$ is replaced by a featureless spectrum.

\begin{figure}[htbp]
  \begin{center}
  \includegraphics[width=0.95\linewidth]{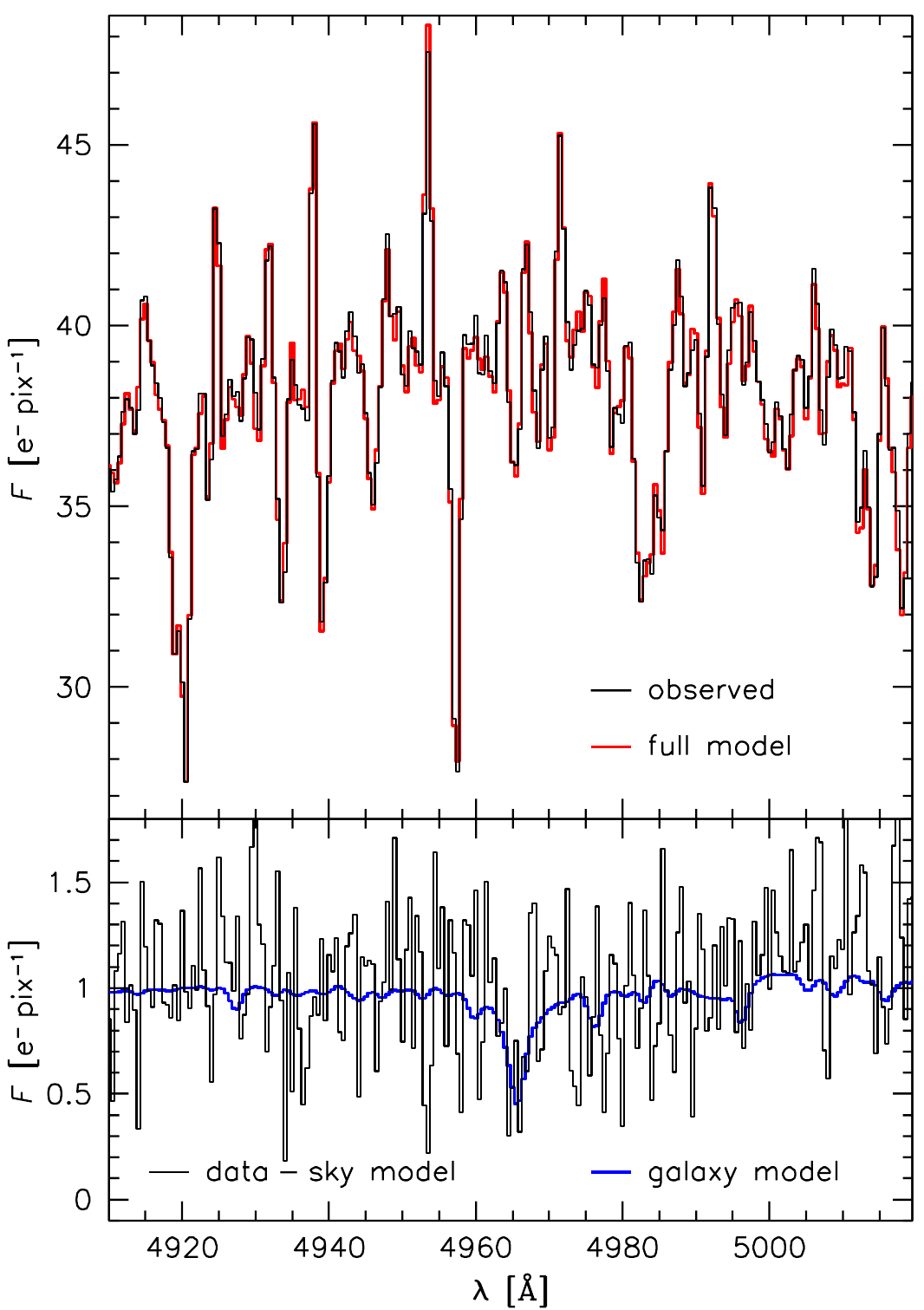}
  \end{center}
\vspace{-0.2cm}
    \caption{
Fit to the sky background in the region around the redshifted H$\beta$
line at $\lambda \approx 4965$\,\AA, for one
of the 34 science exposures (from February 13 2018).
The observed sky spectrum, extracted from the
outer regions of the science frame, is shown in black.
The red line shows the best-fitting full model (Eq.\ 2), which
includes a template for \gal.
The bottom panel shows the residual after subtracting the sky-only model
(that is, the full model
without the galaxy template), as well as the residual
offset calculated in \S\,\ref{res.sec}.
The galaxy template is shown in blue. 
}
\label{skyfit.fig}
\end{figure}

The fit is performed with the {\tt emcee} Markov chain Monte Carlo (MCMC)
sampler ({Foreman-Mackey} {et~al.} 2013), finding the best fitting twelve free parameters
$\alpha_0, ..., \alpha_{11}$ for each of the 34 science cubes.
One hundred walkers are used and 1000 samples are generated. Burn-in
is assumed to occur after 800 samples. Broad, uniform priors
are used. The walkers always converge quickly and
produce well-defined best fit values for all coefficients. The
uncertainties in $\alpha_1, ..., \alpha_{10}$ are
uncorrelated, as expected.
The fits are generally excellent, with residuals consistent with the
expected photon noise. An example is shown in Fig.\ \ref{skyfit.fig},
for a science exposure from the night of February 13 2018. 
The full sky + galaxy model, as given by Eq.\ 2,
is an excellent fit to the data.
For clarity
only the wavelength region around the redshifted H$\beta$ line is shown;
the fit is of similar quality elsewhere.

For each of the 34 science exposures the best-fitting sky model
for that exposure is subtracted from each spatial pixel. The sky model
is given by
\begin{equation}
M_{\rm sky}(\lambda) = M_{\rm full}(\lambda) - \alpha_{11}T_{\rm gal}(\lambda),
\end{equation}
that is, all the sky components of the full model but not its
galaxy template. 
The residual $F-M_{\rm sky}$ is shown in the bottom panel of
Fig.\ \ref{skyfit.fig}, along with the galaxy template.
The H$\beta$ line is detected in the outer
regions of the galaxy in this individual 1800\,s exposure, illustrating
the power of KCWI for studying faint, spatially-extended emission.

\subsubsection{Residual offset}
\label{res.sec}

Our sky subtraction methodology is insensitive to background signal
that shows no variation with wavelength, and also
to signal that is not
present in the blank sky exposures but only in the science cubes.
There is evidence for such signal, as the value of
$\alpha_{11}$, which accounts for all contributions to the science
data that are {\em not} accounted for by the sky templates, varies
between $\alpha_{11}=0.6$ and $\alpha_{11}=3.1$. Inspection of the
H$\beta$ absorption line in the sky-subtracted spectra (see
Fig.\ \ref{skyfit.fig}) indicates that it is not the galaxy flux
that varies,\footnote{Furthermore, it would be difficult to come up
with an explanation why the galaxy flux would vary by a factor
of $\sim 5$, as the
34 science exposures that were retained
were taken under (nearly) photometric
conditions.}  as the absolute absorption (in e$^{-}$\,pix$^{-1}$)
is independent of the value of
$\alpha_{11}$.

\begin{figure}[htbp]
  \begin{center}
  \includegraphics[width=0.95\linewidth]{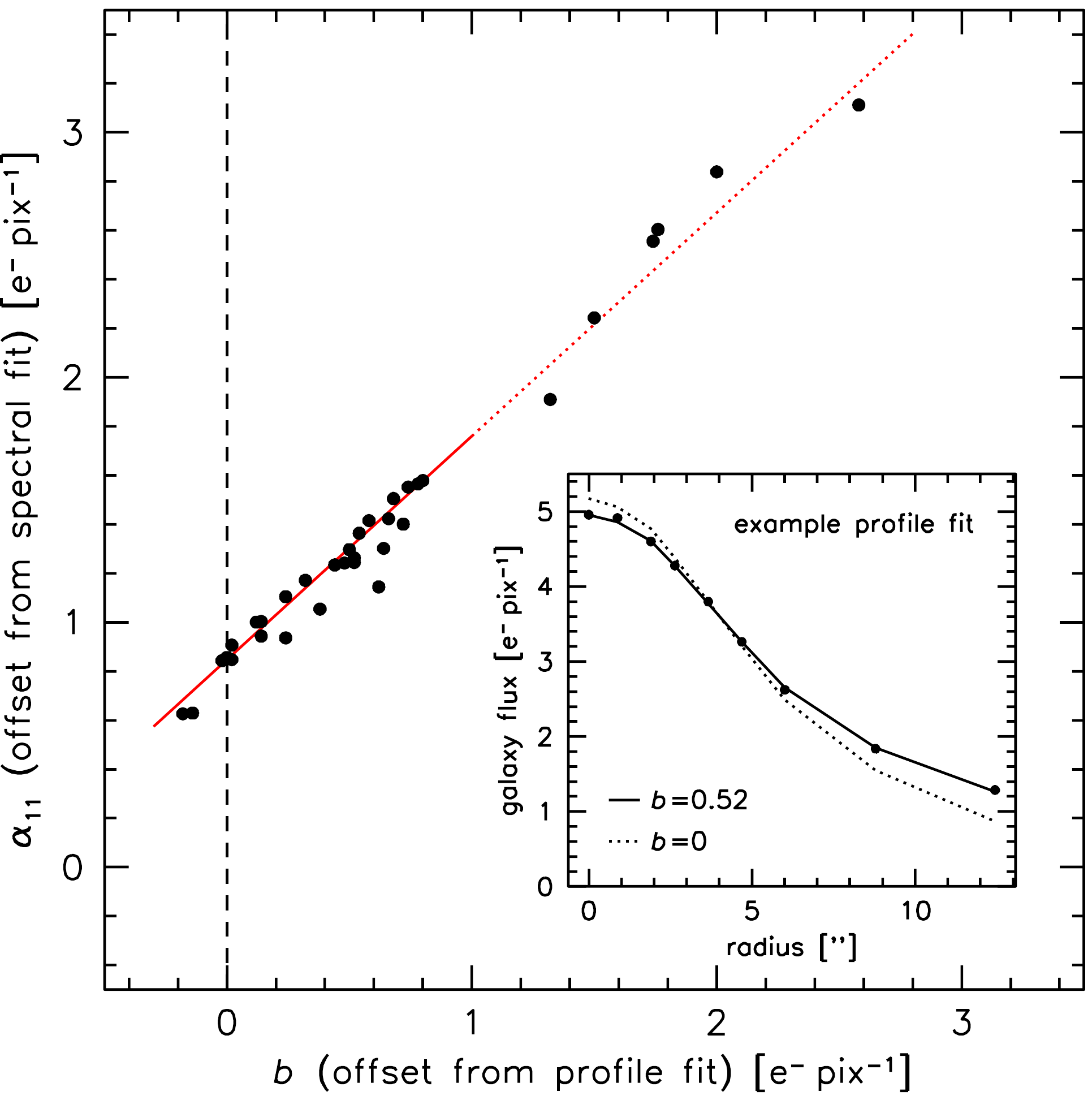}
  \end{center}
\vspace{-0.2cm}
    \caption{
Residual background emission after subtracting the sky model.
The vertical axis is the residual as determined by the {\tt emcee}
fitting of the PCA templates to the science data. The horizontal
axis is the background as determined from fitting the surface
brightness profile of \gal\ with a model that is based on the
HST image of the galaxy. An example of such a fit is shown in the
inset. The two independently-determined values are in very good agreement.
The red line is a least squares fit to the data points. We derive
a contribution of the galaxy flux to $\alpha_{11}$ of 0.85\,e$^{-1}$\,pix$^{-1}$.
}
\label{background.fig}
\end{figure}

The remaining offsets in the sky-subtracted science cubes are
measured in the following way. We measure the average flux
per pixel in elliptical apertures within
the wavelength-collapsed data cubes (the third panel from
left in Fig.\ \ref{hst_kcwi.fig}), carefully masking
contaminating objects. For a perfect sky subtraction
this measurement should correspond to
the surface brightness profile of \gal. We compare these profiles
to the actual surface brightness profile, determined from the
(degraded) HST $V_{606}$ image of \gal\ (second panel in Fig.\ \ref{hst_kcwi.fig}).
We minimize the difference
\begin{equation}
\label{sbfit.eq}
d=\sum_{r}\left[\mu_{\rm KCWI}(r) - (a\,\mu_{\rm HST}(r) + b) \right]^2,
\end{equation}
with $a$ and $b$ free parameters. The fit is done for radii
$r>3\arcsec$
so that variations in the seeing and centering do not influence the
results. The values of $b$ should correspond to
the remaining
background levels in the 34 sky-subtracted science frames. The
process is illustrated for one exposure (the same one as
shown in Fig.\ \ref{skyfit.fig}) in the inset of Fig.\ \ref{background.fig}.
For $b=0$ the scaled HST surface brightness
profile (dotted
line) is not a good fit to the profile measured from the collapsed
data cube (points). The best fit is
obtained for $b=0.52$ (solid line).

In the main panel of Fig.\ \ref{background.fig} the offset that
is derived
from the spectral fit ($\alpha_{11}$) is compared to the offset
as derived from the surface brightness profile fit ($b$).
There is excellent agreement with an offset,
demonstrating that $\alpha_{11}$
indeed represents both residual background and galaxy flux.
We determine the contribution from the galaxy to $\alpha_{11}$
by calculating the
intersection of the relation between $\alpha_{11}$ and
$b$ (red line) with the line $b=0$ (dashed line). We find
$\alpha_{11}(0) = 0.85\pm 0.03$, that is, the galaxy flux
within the aperture that is used for the sky background fit
is 0.85\,e$^{-1}$\,pix$^{-1}$ and
the residual offset that needs to be subtracted from the science
data cubes is $\alpha_{11}-0.85$. We note that this process for determining
the absolute background level ignores a possible color gradient between
the effective wavelength of the $V_{606}$ filter ($0.59\,\mu$m)
and the central wavelength of the spectra ($0.51\,\mu$m). As
shown in Appendix A, the $V_{606}-I_{814}$ color of \gal\ is constant within
the measurement errors at $r>3\arcsec$.

Taking all the results
from this section together, the final, sky- and background-subtracted
data cubes are given by
\begin{equation}
\label{finalsky.eq}
F_{\rm sub}(\lambda) = F(\lambda) - \left( \alpha_0 S_{\rm avg}(\lambda)
+ \sum_{i=1}^{10}
\alpha_i {\rm PCA}_i(\lambda) + \alpha_{11} - 0.85 \right).
\end{equation}
The values of $\alpha_0-\alpha_{11}$ are different for each of the
34 cubes, but
we apply the same sky and background correction for
every spatial pixel. We find no evidence for systematic variations
within the KCWI field of view but this cannot be ruled out.

\subsection{Extraction and combination of spectra}
\label{apertures.sec}

\subsubsection{Definition of apertures}

\begin{figure}[htbp]
  \begin{center}
  \includegraphics[width=0.95\linewidth]{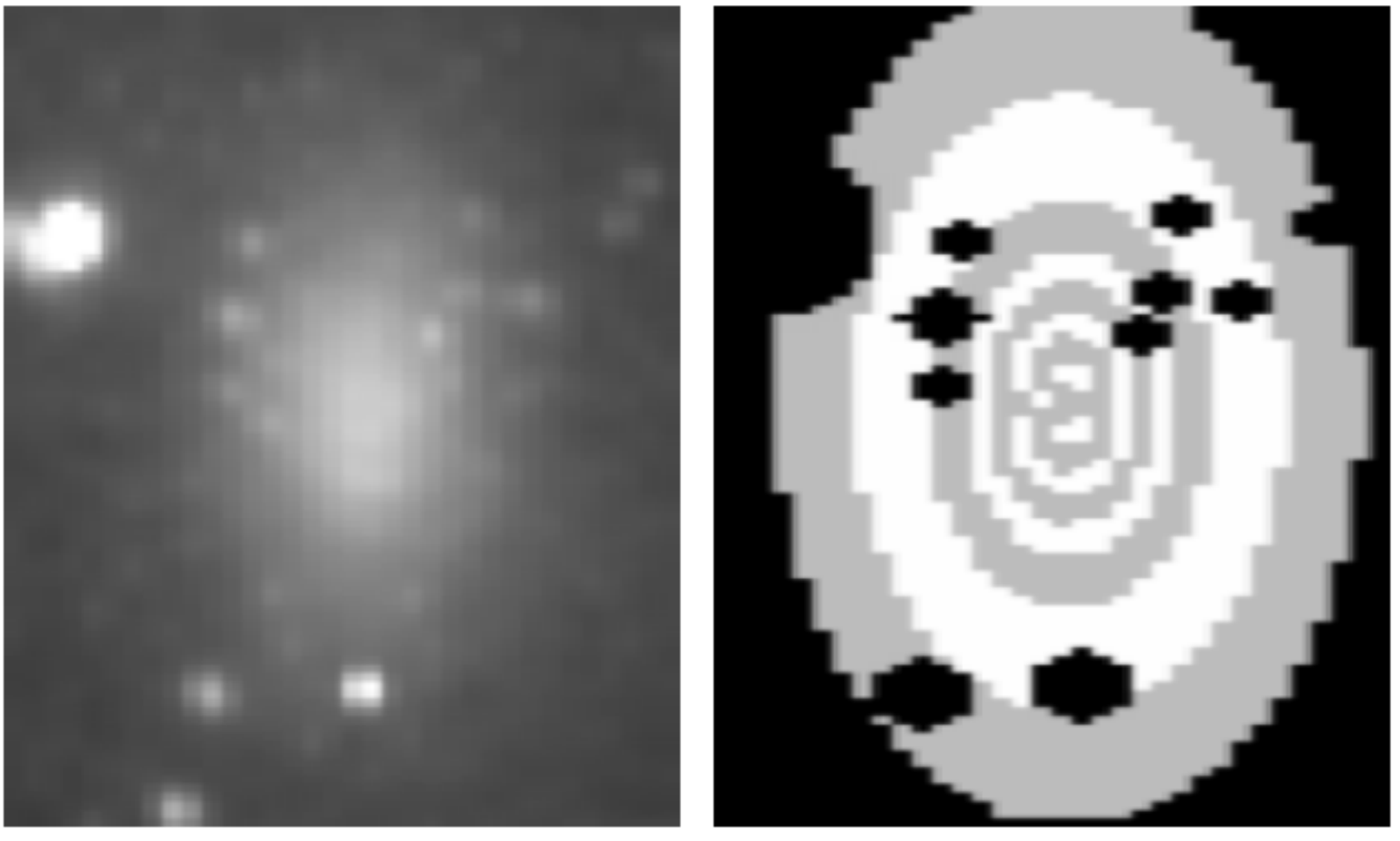}
  \end{center}
\vspace{-0.2cm}
    \caption{
{\em Left:} Degraded HST image of \gal\ (see Fig.\ 2).
{\rm Right:}
Elliptical apertures that are used to determine the kinematics of
\gal\ as a function of radius. Eight apertures are used, from
$r=0\arcsec$ to $r=12\farcs 4$. 
}
\label{apertures.fig}
\end{figure}

One-dimensional spectra are extracted from each of the 34 sky-
and background-subtracted science
cubes, using a variety of spatial apertures. The apertures
that are used in the kinematic modeling are
eight elliptical annuli that follow the isophotes of the galaxy
(see Fig.\ \ref{apertures.fig}).
An approximate square root spacing is used; given the surface brightness
profile of the galaxy this yields an approximately
equal S/N ratio for each of the extracted spectra. The radii,
measured along the major axis of the ellipse, are listed in
Table 2. The axis ratio of the ellipses is 0.69,
and the radius along the major axis can be converted to the circularized
radius through $R_{\rm circ} = 0.83 R_{\rm maj}$.

We also define apertures along the major and minor axis, as well as
masks for contaminating objects and special apertures for particular
objects in the field.
For each aperture and each science exposure a spectrum is obtained
by summing the spectra of all unmasked pixels and dividing the
summed spectrum by the number of unmasked pixels. The number of unmasked
pixels in the outer apertures is not the same for each exposure,
due to the spatial dithering.

\subsubsection{Combination of spectra}
\label{combinespecs.sec}

Before combining the 34 individual spectra for each aperture the 
barycentric velocity
correction needs to be applied, as the data were taken over a long time
period.  The correction ranges from $22$\,\kms\ for
the January data to $-18$\,\kms\ in April, which means the variation is of the same order as
the instrumental resolution and the central velocity dispersion of the galaxy.
To minimize interpolation-induced smoothing the
spectra are resampled onto a $2\times$ finer grid with
d$\lambda=0.25$\,\AA\ in this step. In this resampling step a small wavelength calibration
correction
is applied, as derived from the fit of the Solar spectrum to PC1 (see \S\,\ref{solar.sec}).
Omitting this correction does not lead to discernable changes in the results.

\begin{figure*}[htbp]
  \begin{center}
  \includegraphics[width=0.95\linewidth]{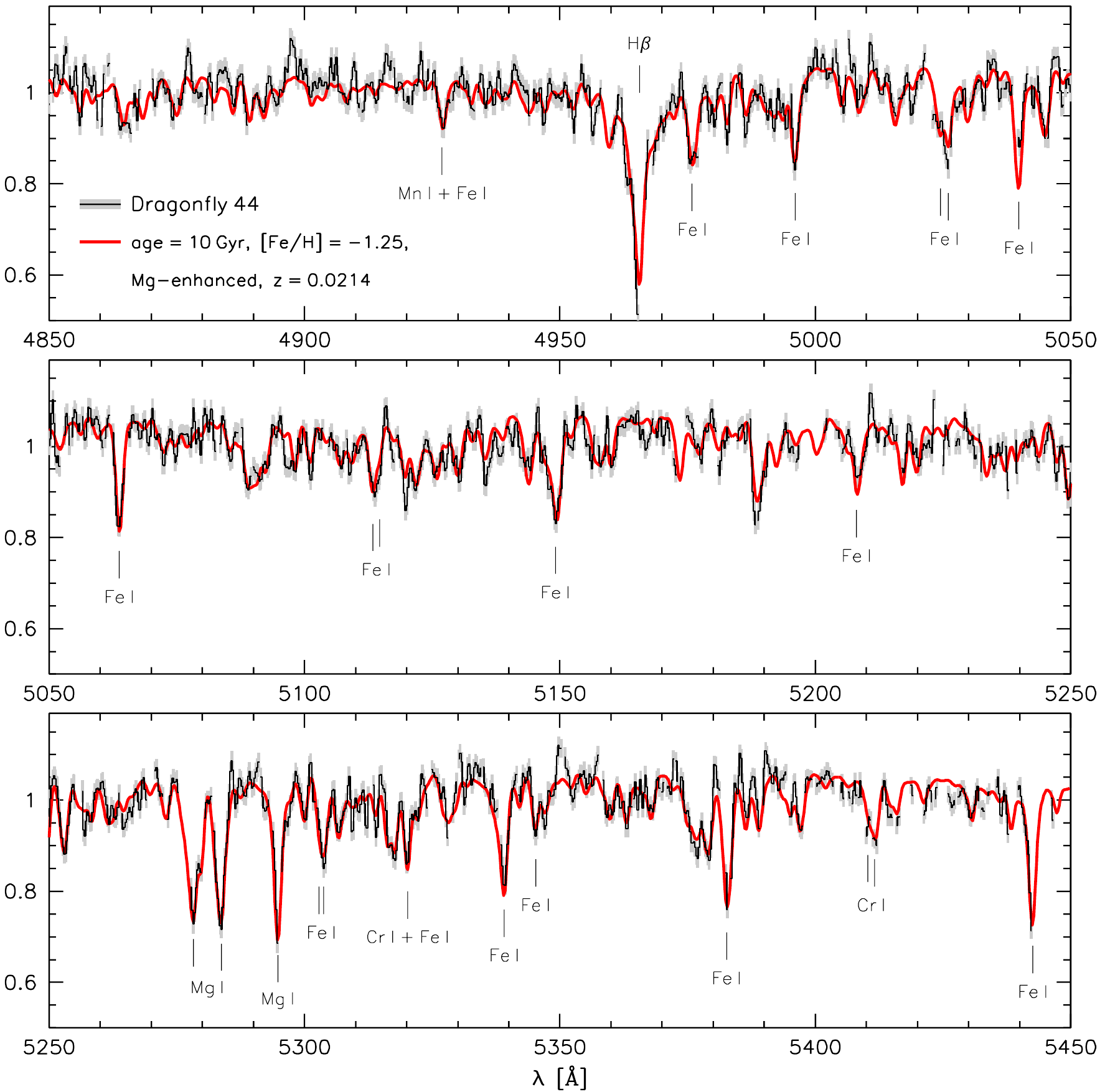}
  \end{center}
\vspace{-0.2cm}
    \caption{
S/N-optimized integrated 17\,hr KCWI
spectrum of \gal\ (black, with $1\sigma$
uncertainties in grey). The median S/N ratio is 48\,pix$^{-1}$,
or 96\,\AA$^{-1}$. The synthetic template spectrum that is used to measure the kinematics
is shown in red. 
}
\label{optimal.fig}
\end{figure*}

The individual shifted spectra are then combined:
\begin{equation}
F_{\rm avg}(\lambda) = \frac{\sum_{i=1}^{34} n_i w_i F_{{\rm sub}, i}(\lambda)}
{\sum_{i=1}^{34} n_i w_i},
\end{equation}
with $n_i$ the number of unmasked pixels of exposure $i$ and $w_i$ the weight.
These weights are defined as
\begin{equation}
w_i = \frac{a_i^2}{m_i} \Big\langle \frac{a_i^2}{m_i}
\Big \rangle^{-1}_i,
\end{equation}
with $a$ the galaxy scaling parameter from Eq.\ \ref{sbfit.eq}, $m$
the wavelength-averaged flux 
of model $M$ (Eq.\ 2), and angular brackets indicating the
mean. The weights vary between 0.79 and 1.29
with an rms of 0.12.

The uncertainties in the averaged spectra are determined from a combination of the
averaged sky spectrum and the residuals from a fit to a stellar population synthesis
model:
\begin{equation}
e(F_{\rm avg})(\lambda) = \frac{s_{\rm bi}}{m_{\rm avg}} \sqrt{M_{\rm avg}(\lambda)}.
\end{equation}
Here $M_{\rm avg}$ is calculated in an analogous way to $F_{\rm avg}$,
$m_{\rm avg}$ is the wavelength-averaged mean of $M_{\rm avg}$, and $s_{\rm bi}$
is the biweight scatter ({Beers}, {Flynn}, \& {Gebhardt} 1990) in the residuals from a fit to a stellar
population synthesis model (see \S\,\ref{kin.sec}).
This approach ensures that the fits in \S\,\ref{kin.sec} have acceptable $\chi^2$ and
that the ensuing uncertainties in the fit parameters are properly normalized.
Finally, we mask pixels with residuals that exceed $2.5\times$ the expected
error. These are invariably associated with the strongest sky lines, which are
imperfectly modeled with the PCA analysis. In all apertures the fraction of masked pixels
in the wavelength range $4850$\,\AA\,$<\lambda< 5450$\,\AA\ is 2--3\,\%.

\subsection{Optimal extraction}

In addition to the aperture spectra described above we create a
combined spectrum of the entire galaxy
that maximizes the S/N ratio:
\begin{equation}
F_{\rm opt}(\lambda) = \frac{\sum_{i=1}^8 w_i F_{{\rm avg,aper}\,\,i}
(\lambda)}{\sum_{i=1}^8 w_i},
\end{equation}
with the sums over the 8 elliptical apertures and the weight given
by
\begin{equation}
w_i = \frac{f_{{\rm avg,aper}\,\,i}}{e_{{\rm
avg,aper}\,\,i}}.
\end{equation}
Here $f_{{\rm avg,aper}\,\,i}$ is the wavelength-averaged signal and
$e_{{\rm avg,aper}\,\,i}$ is the wavelength-averaged error in aperture $i$.

This spectrum provides the best
constraints on the average stellar population of \gal\
(A.~Villaume et al., in preparation). In the present study
it is used to select the
template that is
fit to the data in our kinematic modeling (\S\,\ref{template.sec}) and to constrain
the kinematic line profile (\S\,\ref{h3h4.sec}).
It is shown in Fig.\ \ref{optimal.fig}, along with the best-fitting
model from \S\,\ref{kin.sec} for reference.
The median S/N ratio is 48 per 0.25\,\AA\ pixel, equivalent to
96\,\AA$^{-1}$.

\section{Measurements of Kinematics}
\label{kin.sec}

\subsection{Modeling of the spectral resolution}
\label{solar.sec}

As discussed below we use a stellar population synthesis model as a template
to measure the kinematics of \gal. This requires that the template and the
data have the same resolution. We therefore need
to accurately characterize the (wavelength-dependent)
resolution that is delivered by the instrument.

\subsubsection{Fitting the solar spectrum}

The instrumental resolution is
typically determined from the widths of emission lines in
arc lamp exposures; however, both the light path and
the data handling of the calibration
lamps are different from the science data. The
data reduction process of the science data involves the combination of long
exposures over
many nights and this is likely to impact
the effective spectral resolution. Furthermore, the kinematics are measured
from template fits to absorption line spectra rather than from the fits
of Gaussians to individual emission lines.

Ideally, the instrumental resolution is measured directly from the science data,
for example by using higher resolution observations of the same objects as templates
(see {van Dokkum} {et~al.} 2017b). In our case, we make use of the fact that one of the
eigenspectra of the sky variation (PC1; see Fig.\ \ref{pca.fig})
comprises scattered and reflected sunlight. We fit PC1 with a high resolution
solar spectrum obtained from the BAse de donn\'ees Solaire Sol
(BASS2000\footnote{\url{http://bass2000.obspm.fr/solar\_spect.php}}), in small
wavelength intervals. Both the model and the data were divided
by a polynomial of order $(\Delta \lambda/100)+1$. Free parameters in the
fit are the radial velocity, the velocity dispersion, and an additive constant.
The instrumental line profile is held fixed, using
$h_3^+=-0.005$ and $h_4^+=-0.094$ (see below). 
The fit is done using the {\tt emcee}-based code described in {van Dokkum} {et~al.} (2016).

\begin{figure*}[htbp]
  \begin{center}
  \includegraphics[width=0.75\linewidth]{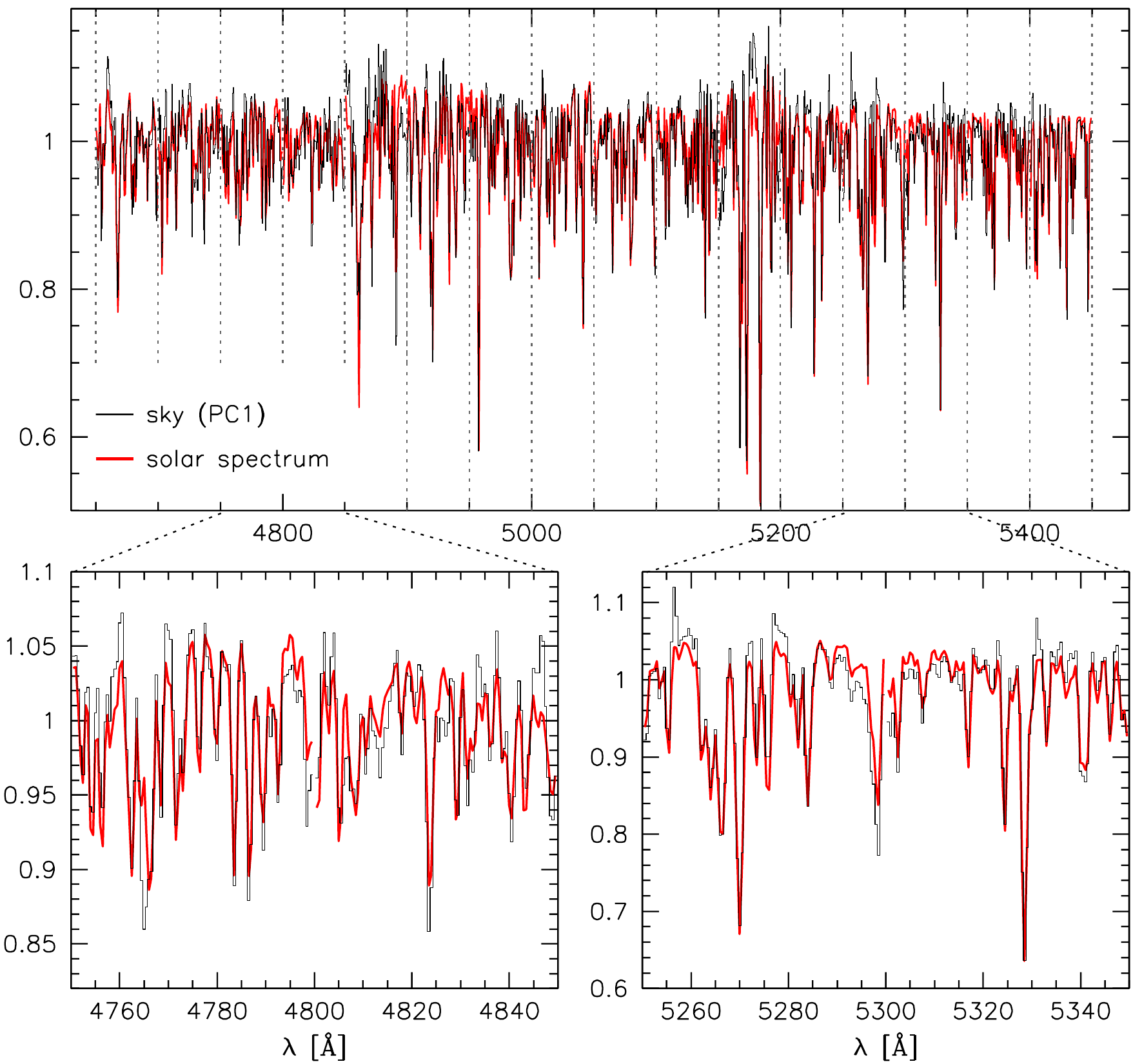}
  \end{center}
\vspace{-0.2cm}
    \caption{
Continuum-normalized
first principal component (PC1)
from Fig.\ \ref{pca.fig}. The red line is a high resolution
solar spectrum, fit in small wavelength intervals to PC1 to determine the
wavelength-dependent spectral resolution and the accuracy of the
wavelength calibration.
}
\label{solar.fig}
\end{figure*}

\begin{figure*}[htbp]
  \begin{center}
  \includegraphics[width=0.8\linewidth]{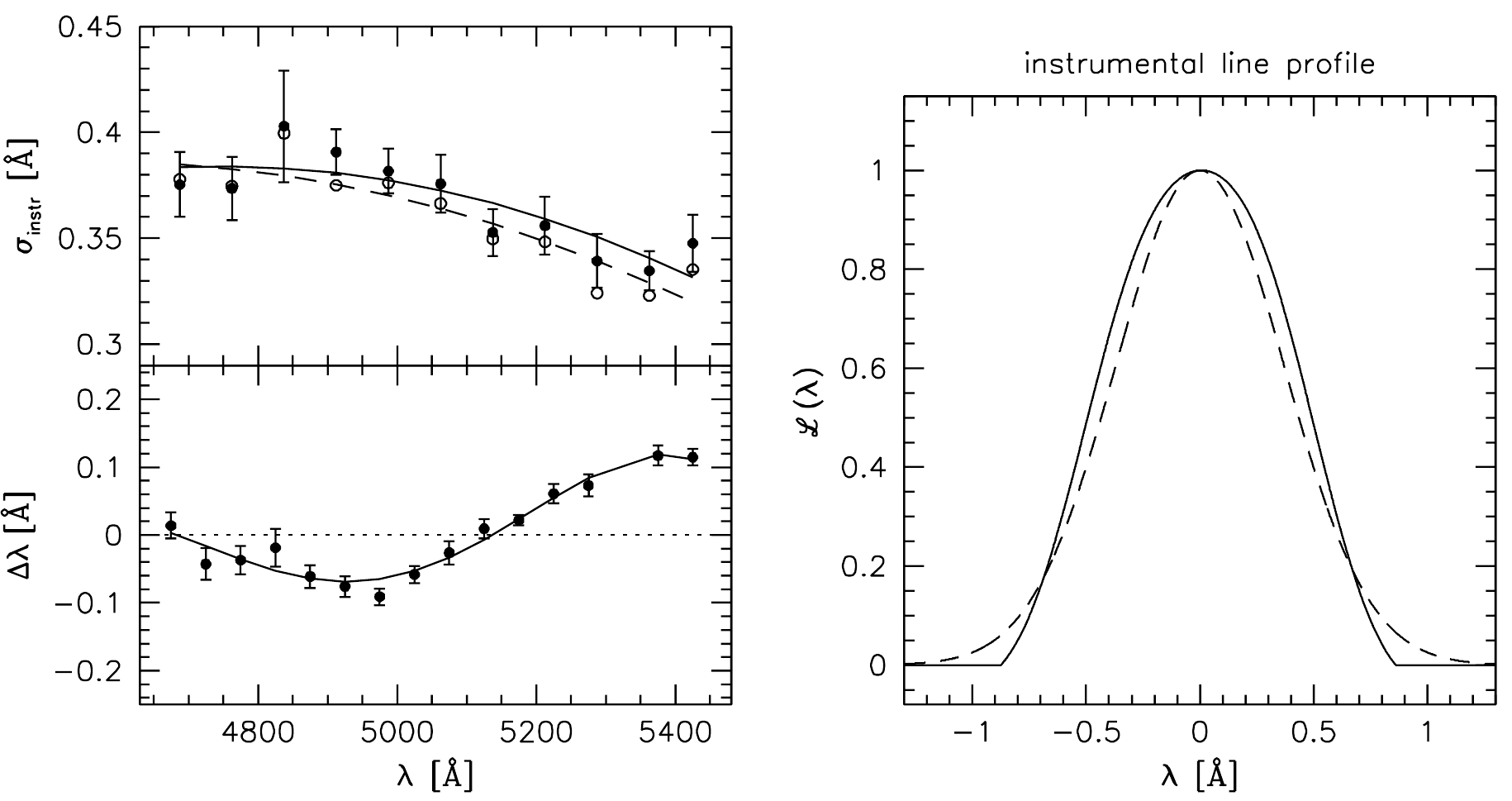}
  \end{center}
\vspace{-0.2cm}
    \caption{
Wavelength-dependent spectral resolution, as determined from PC1.
{\em Top left:} Second moment of the instrumental resolution for a Gaussian
(open circles; dashed line) and non-Gaussian (solid circles; solid line)
line profile. {\em Bottom left:} Error in the wavelength calibration.
{\em Right:} Average instrumental line profile (solid line), as determined from
fitting the asymmetry and skewness in three wavelength intervals, compared
to a Gaussian (dashed line). In the main analysis the measured non-Gaussian line profile
was used, as well as the wavelength calibration correction. These choices do not
affect the final results.
}
\label{resolution.fig}
\end{figure*}

The best fits
are shown in Fig.\ \ref{solar.fig}. The correspondence between PC1 and the solar
spectrum is remarkably good, as illustrated by the insets.
The resulting instrumental resolution is shown by the solid
points in the top left panel
of Fig.\ \ref{resolution.fig}.
The line is the best-fitting relation, of the form
\begin{equation}
\label{resol.eq}
\sigma_{\rm instr}(\lambda) = 0.377 -5.79\times 10^{-5} \lambda_{5000} - 1.144
\times 10^{-7} \lambda_{5000}^2,
\end{equation}
with $\lambda_{5000} = \lambda - 5000$ and $\sigma_{\rm instr}$ the second moment
of the instrumental resolution in units of \AA.
This corresponds to $\sigma_{\rm instr}=24.0$\,\kms\ at $\lambda=4800$\,\AA\ and
$\sigma_{\rm instr}=18.6$\,\kms\ at $\lambda=5400$\,\AA, and $R\approx 5,600$
at $\lambda=5000$\,\AA.
Forcing the instrumental profile to be Gaussian ($h_3^+ = h_4^+ = 0$) produces
very similar results, as shown by the open symbols and dashed line.

The relative velocity shift with wavelength,
expressed in \AA, is shown in the bottom left panel of Fig.\
\ref{solar.fig}. 
We find a small but systematic wavelength calibration error, with peak-to-peak
variation of $\pm 0.1$\,\AA\ ($\pm 6$\,\kms). The rms variation is $0.06$\,\AA\
(4\,\kms). A fourth-order polynomial fit to this variation (indicated by the
solid line) is included in the resampling of the science data
(see \S\,\ref{combinespecs.sec}).

\subsubsection{Instrumental line profile}

A major uncertainty in
measuring mass profiles from kinematic data is the degree of anisotropy
in the velocity distribution,
and this can, in principle, be constrained by deviations from a Gaussian profile:
flat-topped profiles indicate tangential anisotropy, peaked profiles
radial anisotropy (see, e.g., {Bender}, {Saglia}, \& {Gerhard} 1994; {Thomas} {et~al.} 2007; {Amorisco} \& {Evans} 2012).
However, this relies on an excellent characterization of the {\em instrumental}
line profile, as well as a very high S/N ratio and adequate
control of systematics such as the wavelength calibration.

To characterize the instrumental line profile we follow
common practice and parameterize deviations from a Gaussian profile with the
(asymmetric) $h_3$ and (symmetric) $h_4$ components of a Gauss-Hermite
expansion
 ({van der Marel} \& {Franx} 1993; {Cappellari} {et~al.} 2007).
The line of sight velocity
distribution is then parameterized by
\begin{equation}
L(y) = \frac{\exp(-y^2/2)}{\sigma\sqrt{2\pi}} \left[ 1+
h_3 \frac{y(2y^2-3)}{\sqrt{3}} + h_4\frac{4(y^2-3)y^2 + 3}{\sqrt{24}}\right],
\end{equation}
with $y = (v-V)/\sigma$ (see, e.g., {Cappellari} 2017).
As the line profile cannot be negative, we impose the modification
\begin{equation}
L'(y) = 
\begin{cases}
  L(y) & \text{if $L(y)>0$} \\
  0  & \text{otherwise}
\end{cases}.
\end{equation}
We use $h_3^+$, $h_4^+$ to specify the components of this modified profile, and
include these in the {\tt emcee}-based dispersion fitting code.

We find that the line profile is flat-topped, with negative $h_4^+$,
as expected for the slitwidth-limited resolution provided by
the medium slicer  (see, e.g., {Casini} \& {de Wijn} 2014).
There is no clear trend with wavelength, and the average values for the
Gauss-Hermite components in three wavelength regions
are $\langle h_3^+\rangle =-0.005 \pm 0.005$ and $\langle h_4^+\rangle =-0.094 \pm 0.014$.
The corresponding profile is shown in the right panel of Fig.\ \ref{resolution.fig}.
We use this line profile in the template construction in \S\,\ref{template.sec}.

\subsection{Template construction}
\label{template.sec}

The kinematics are measured by fitting a template spectrum to the observed spectra.
The template is one of a set of synthetic stellar population synthesis models that
have a native resolution of $R=10,000$ and are based on the same set of libraries as
discussed in \S\,\ref{expect_spec.sec}. These model spectra
are convolved to the resolution of the \gal\ spectra. 
This convolution takes the
native resolution of the templates into account, as well as the
line profile and the wavelength dependence of the KCWI resolution (as determined
in \S\,\ref{solar.sec}). We note that the line profile
of the synthetic templates is (exactly) Gaussian.

\begin{figure}[htbp]
  \begin{center}
  \includegraphics[width=0.95\linewidth]{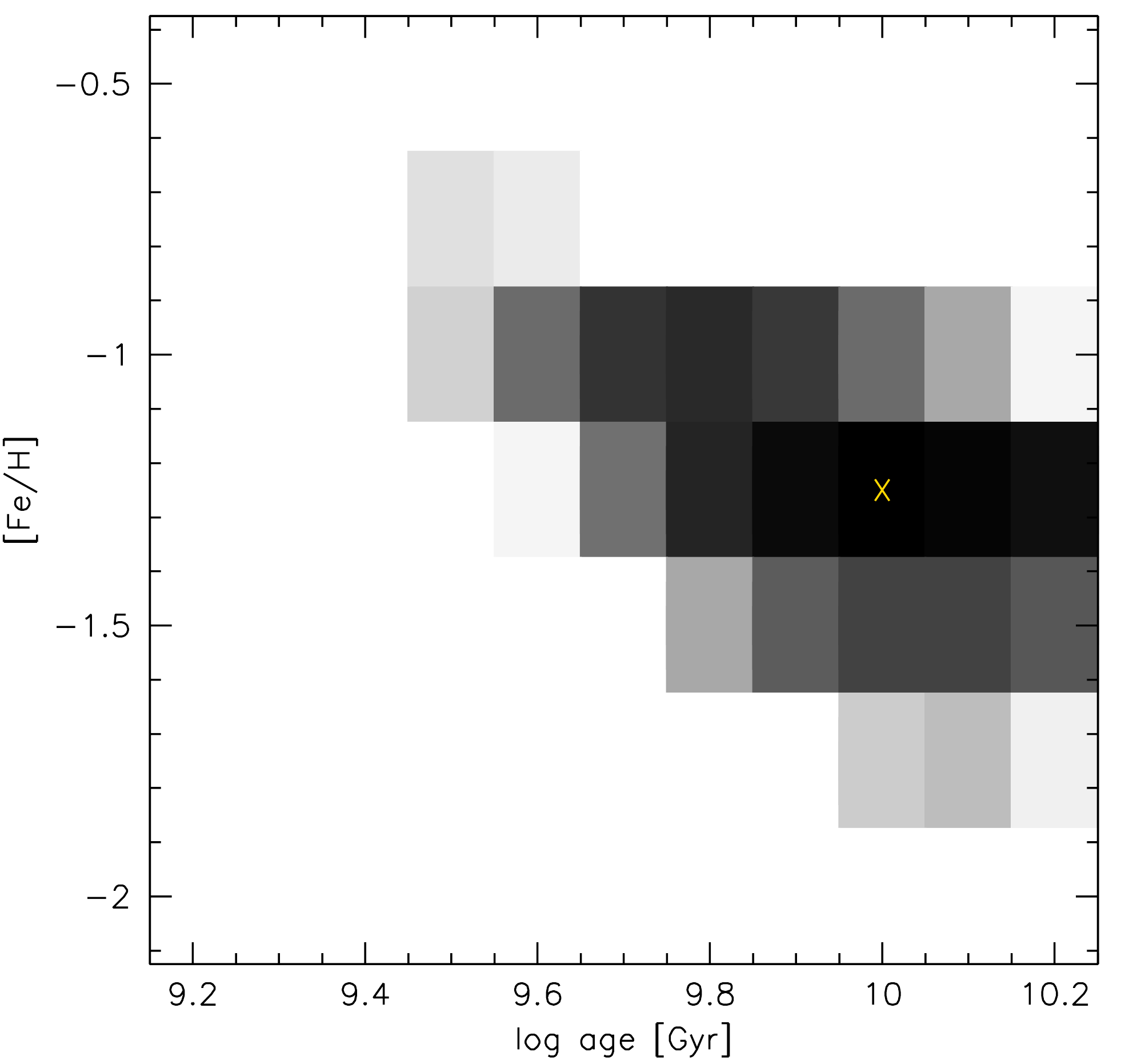}
  \end{center}
\vspace{-0.2cm}
    \caption{
Determination of the age and metallicity of the template spectrum. High resolution
stellar population synthesis models, convolved to match the instrumental resolution,
are fit to the optimally-extracted KCWI spectrum of \gal.  Darker regions
indicate a higher log likelihood. The best fit (denoted by the yellow X)
is obtained for an age of 10\,Gyr and
[Fe/H]$=-1.25$.
}
\label{temp_chisq.fig}
\end{figure}

The best-fitting template is determined by fitting models with a range of
discrete ages
and metallicities to the optimally-extracted spectrum shown in Fig.\
\ref{optimal.fig} and determining the relative likelihood. The results are shown
in Fig.\ \ref{temp_chisq.fig}, with the grey level indicating the relative log likelihood.
The distribution of the grey points follows the well-known age-metallicity degeneracy,
and the best fit
is obtained for an age of 10\,Gyr and a metallicity [Fe/H]$=-1.25$. This result
is in good agreement with the previous measurement by {Gu} {et~al.} (2018), who
found an age of $8.9^{+4.3}_{-3.3}$ Gyr and [Fe/H]$=-1.3^{+0.4}_{-0.4}$ for \gal\
from deep MaNGA spectroscopy.
Quantitative constraints on the (spatially-resolved)
stellar population of \gal, derived from more flexible lower resolution model
fits to the KCWI data, will be presented in A.~Villaume et al., in preparation.

The high resolution models we use have solar abundance ratios, whereas stellar populations
with ages of $\sim 10$\,Gyr are often $\alpha$-enhanced.
We model an enhanced Mg abundance by artificially increasing the depth of
the Mg triplet lines, using high resolution
integrated KCWI spectra of the old and metal poor Milky Way
globular clusters M3 and M13. These data were obtained
as templates to measure the velocity dispersion of the UDG NGC\,1052-DF2
(see {Danieli} {et~al.} 2019).
Here they are used to slightly increase (by $\approx 5$\,\%)
the depth of the Mg lines in the template spectrum,
matching them to the observed lines in M3 and M13.
We tested that this enhancement has a negligible effect on the final dispersions.
The final template is, then, a synthetic stellar population synthesis model with
an age of 10\,Gyr, [Fe/H]$=-1.25$, and [Mg/Fe]$\approx 0.3$, convolved to match the
wavelength-dependent non-Gaussian line profile of KCWI.

\begin{figure}[htbp]
  \begin{center}
  \includegraphics[width=0.95\linewidth]{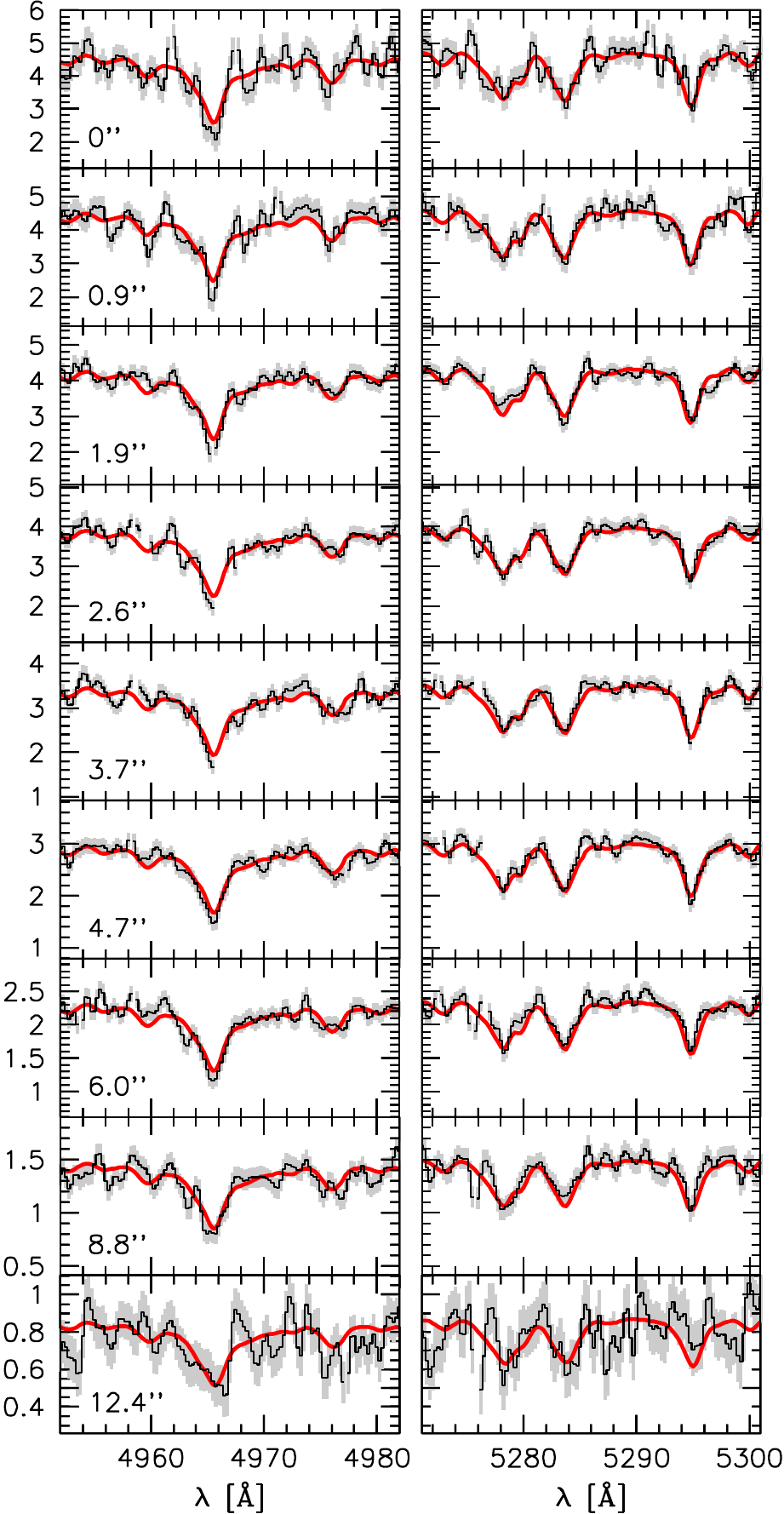}
  \end{center}
\vspace{-0.2cm}
    \caption{
Fits in elliptical annuli, from $r=0\arcsec$ to $r=12\farcs 4$. The fits
are done over the same wavelength interval as in Fig.\ \ref{optimal.fig},
but for clarity only the regions around H$\beta$ (left) and the Mg
triplet (right) are shown. The units are e$^{-}$\,pix$^{-1}$.
}
\label{radial.fig}
\end{figure}

\begin{figure*}[htbp]
  \begin{center}
  \includegraphics[width=0.85\linewidth]{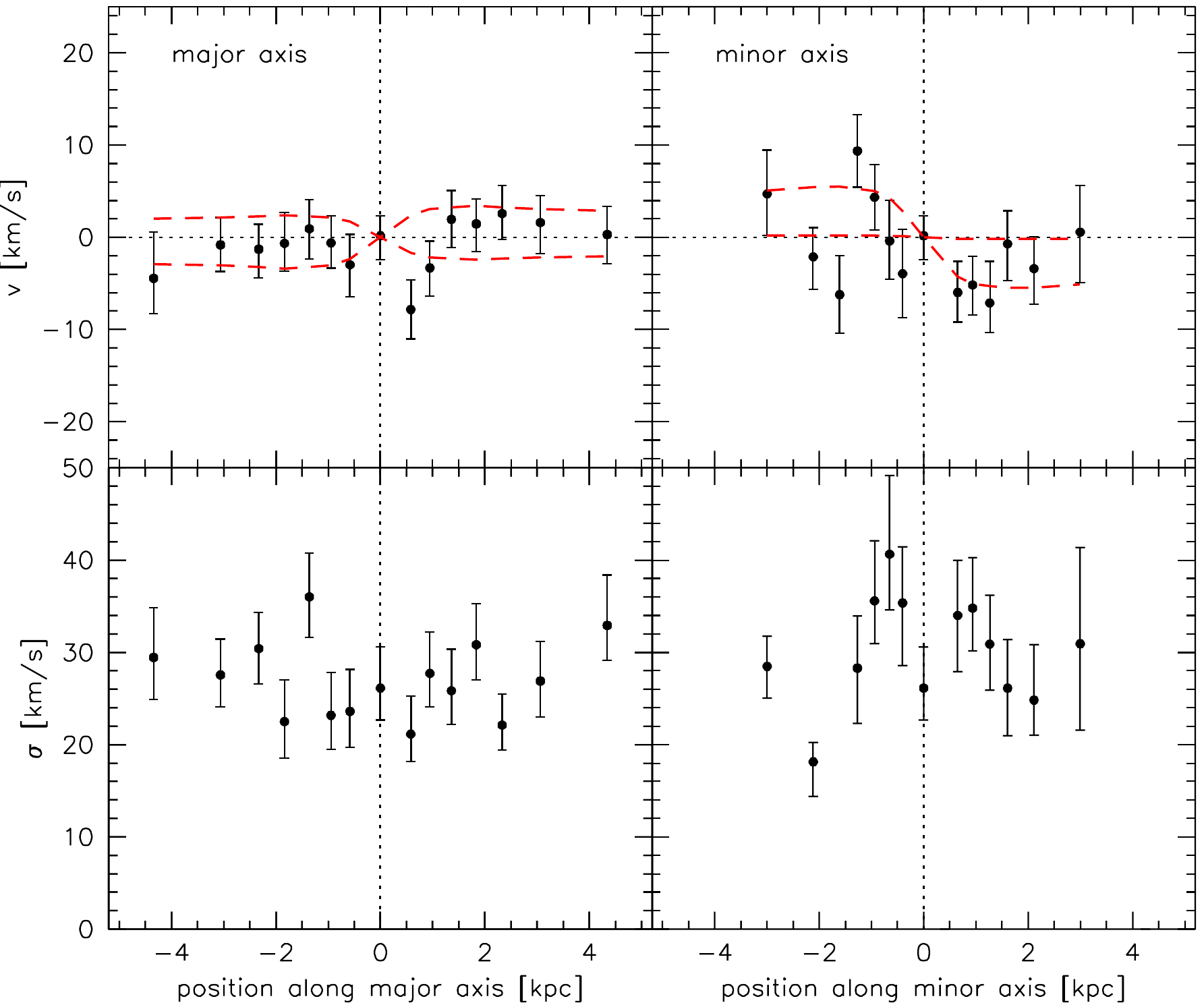}
  \end{center}
\vspace{-0.2cm}
    \caption{
Major and minor axis kinematics of \gal. Velocity profiles are shown in the top
panels and dispersion profiles in the bottom panels. The galaxy does not
show evidence for rotation;  the red dashed curves
indicate 90\,\% upper limits on the rotation velocity.
}
\label{axes.fig}
\end{figure*}

\subsection{Velocity and velocity dispersion measurements}
\label{fitting.sec}

The kinematic fits
are done with the MCMC methodology described briefly above and
more extensively in {van Dokkum} {et~al.} (2016). The template and data were continuum
filtered with
a polynomial of order $\Delta \lambda/100+1$. The template normalization,
velocity, $\sigma$, $h_3$, and $h_4$ are fit parameters. We also
include two optional parameters $c_1$, $c_2$ to describe any remaining wavelength calibration
errors. The template is sampled onto a wavelength grid defined as
\begin{equation}
\lambda_z' = \lambda_z + c_1 (\lambda_z - \langle \lambda_z \rangle) + c_2 (\lambda_z - \langle \lambda_z
\rangle)^2,
\end{equation}
with $\lambda_z = \lambda \times (1+z)$. For our default measurements
$c_1=c_2 = 0$. As discussed above,
in the following
``velocity dispersions'' are actually second moments of the velocity distribution
$L(y)$.  
We use 100  walkers and 1000 samples, with burn-in assumed to occur
after 800. The fits are well-behaved, and provide stable and converged minima.
We fit all of the apertures described in \S\,\ref{apertures.sec}.

In Fig.\ \ref{radial.fig} we show the best fits for
the nine elliptical apertures from Fig.\ \ref{apertures.fig}, with a focus on
the region around the H$\beta$ line and the Mg triplet. Qualitatively, the
fits are excellent, with the red model generally within the grey band around
the data. This is quantified by calculating the residuals from the model fits
and comparing these to the expected errors. For all apertures we find that
the biweight scatter in the residuals corresponds to the median error within
$10$\,\%, as expected. 

We assess the importance of systematic errors by varying the analysis.
Neglecting to correct the Mg strength to match globular clusters leads to a change
in derived dispersions of $\sim 1$\,\kms. Allowing an additive offset or linear
combinations of multiple templates leads to $\sim 5$\,\kms\ changes in the
derived dispersions. We note, however, that these should not
be considered free parameters; the surface brightness profile of the HST image
of \gal\ sets the overall background level, and the template corresponds
to the best stellar population synthesis fit (with small errors) to a
smoothed version of the data. Changing the wavelength region of the fit or
forcing $h_3=h_4=0$  has a 1--2\,\kms effect. 
Splitting the spectra in three equal-length wavelength regions also leads to 1--2\,\kms\
effects. Not applying the wavelength calibration correction function (the fit
in the bottom left panel of Fig.\ \ref{resolution.fig}), or fitting for $c_1$ and
$c_2$,
has a negligible effect on the dispersions and also on the $h_3$ and $h_4$ parameters.
Allowing $c_1$ and $c_2$ to be free does have an effect on the derived velocities;
these show larger scatter with larger uncertainties, as expected.

\section{Kinematics of Dragonfly~44}

\subsection{Major and minor axis kinematics}

We first consider whether the galaxy is supported by rotation, as might be expected
in some UDG formation models (e.g., {Amorisco} \& {Loeb} 2016). 
The kinematics along the major and minor axis are shown in Fig.\ \ref{axes.fig}. They are
measured in wedges that grow from $2\arcsec$ width in the center to $\approx 10\arcsec$ at
the largest distances from the center.
The rotation velocity, with respect to the mean, is shown in the top panels, and the
velocity dispersion is shown in the bottom panels. The major and minor axis profiles
are shown separately. There is no evidence for rotation. We determine the maximum
rotation speed by fitting the normalization of a model rotation
curve to the velocity data. The model is the best fitting
Jeans model to the rotationally-supported dE galaxy NGC\,147, as derived by
{Geha} {et~al.} (2010). NGC\,147 is a satellite
of M31 with a similar stellar mass as \gal. This is an ad hoc
way of generating a plausible rotation curve shape;
our results are not sensitive to the precise
form of the model.

The best-fitting maximum rotation velocities are $V_{\rm max}=1$\,\kms\ and $V_{\rm max}=
3$\,\kms\ for the
major and minor axis. Both values are consistent with zero, and the 90\,\%
upper limits are $V_{\rm max}<3.4$\,\kms\ and $V_{\rm max}<5.5$\,\kms\ respectively.
The mean velocity dispersion is 27\,\kms\ for the major axis and 32\,\kms\ for
the minor axis, which implies $V_{\rm max}/\langle \sigma\rangle < 0.12$ (major axis)
and $V_{\rm max}/\langle \sigma\rangle<0.17$ (minor axis), with 90\,\% confidence.
We note that a physically better-motivated measure of rotational support is
$\left(\langle V^2\rangle/\langle \sigma^2\rangle\right)^{0.5}$, that is, the rms of the locally-measured
ratios (see Binney 2005; Cappellari et al.\ 2007). However, the S/N ratio in individual spatial
bins is not sufficiently high to measure this quantity reliably from our data.

\begin{figure}[htbp]
  \begin{center}
  \includegraphics[width=0.95\linewidth]{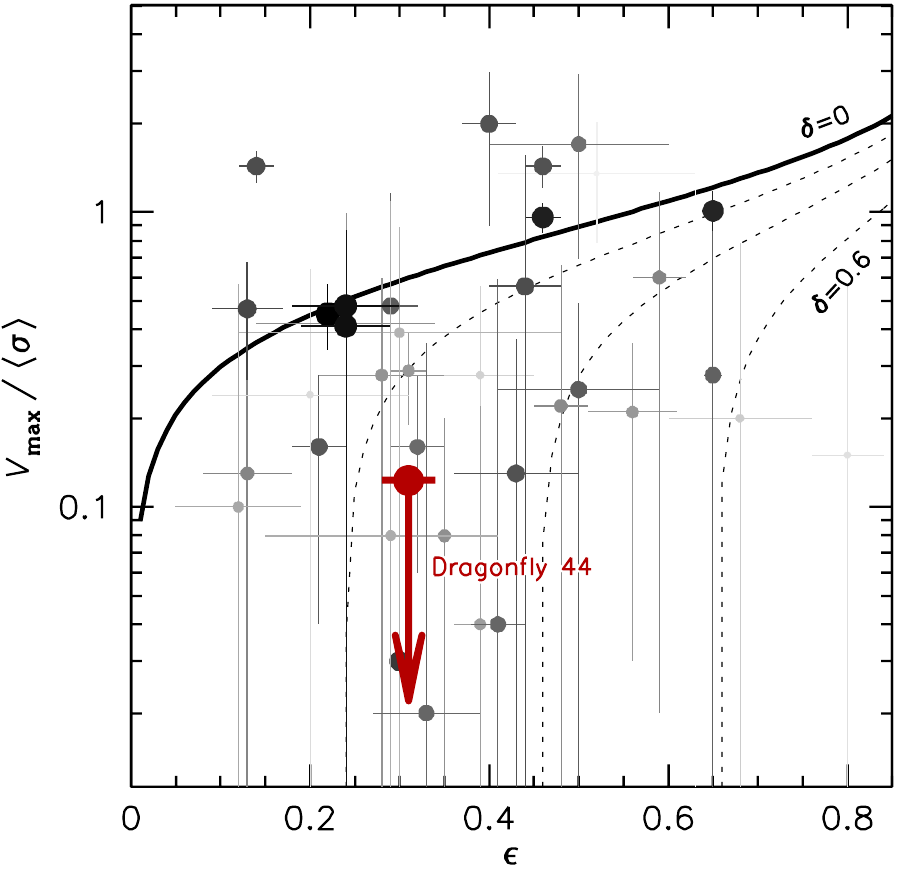}
  \end{center}
\vspace{-0.2cm}
    \caption{
Relation between $V_{\rm max}/\langle \sigma\rangle$ along the major
axis and ellipticity $\epsilon$.
The solid line is for oblate spheroids with no anisotropy; dotted lines are for
increasing anisotropy. Data points are from the compilation of nearby dwarf galaxies
of {Wheeler} {et~al.} (2017), with the grey level and size of the symbol proportional
to $\log(M_{\rm stars})$. The arrow is the 90\,\% upper limit for \gal.
\gal\ is not supported by rotation, and may have a low value of $V_{\rm max}/\langle
\sigma\rangle$ for its stellar mass.
}
\label{vsig.fig}
\end{figure}

The limit on $V_{\rm max}/\langle \sigma\rangle$ along the major axis of \gal, combined with
its axis ratio of $b/a=0.69$, means that the galaxy is not rotationally-supported. 
In Fig.\ \ref{vsig.fig} \gal\ is placed on the well-known {Binney} (1978)
diagram of
$V_{\rm max}/\langle\sigma\rangle$ 
versus observed ellipticity ($\epsilon = 1 - b/a$). The solid line is
for edge-on oblate spheroids with no anisotropy (see the discussion in \S\,6.1
of Cappellari et al.\ 2007). In such models
$V_{\rm max}/\langle\sigma\rangle = 0.6$ for $\epsilon=0.31$,
an order of magnitude higher than the upper limit for \gal.
Dotted lines are for increasing anisotropy, here parameterized with $\delta = (2\beta -\gamma)/
(2-\gamma)$ (see, e.g., Eq.\ 4--7 in {Cappellari} {et~al.} 2007). 
The other data points in Fig.\ \ref{vsig.fig} are dwarf galaxies in and near the Local
Group, taken from the compilation by {Wheeler} {et~al.} (2017).
The most straightforward interpretation of the distribution of the Local Group
dwarfs is that they are supported by random motions rather than rotation (see
Wheeler et al. 2017), and that the observed shapes are due to anisotropy in
the velocity dispersion tensor.
We note that these galaxies
typically have low Sersic indices, similar to \gal.
The grey level and size of the symbol indicate the stellar masses of the galaxies,
which range from $10^4-10^8$\,\msun. 
With  $M_{\rm stars}
\approx 3\times 10^8$\,\msun\ ({van Dokkum} {et~al.} 2016)
the stellar mass
of \gal\ is just above the highest mass
galaxy in the {Wheeler} {et~al.} (2017) sample.
Its $V_{\rm max}/\langle \sigma\rangle$
is at the low end of the distribution of dwarf galaxies, particularly
when compared to more massive dwarfs which tend to have slightly more rotational
support (as also noted in {Wheeler} {et~al.} 2017).

Finally, we note several caveats.
First, the $V/\sigma$ diagnostic diagram has
mostly been applied to luminous early-type galaxies, which are baryon-dominated
within the effective radius. Second, the dynamics of galaxies
can be quite complex. A specific example  is the early-type galaxy NGC\,4550, whose
relatively flat rotation curve and radially-increasing velocity dispersion profile
(Rix et al.\ 1992) are similar to what is seen in \gal\ (see \S\,\ref{sigprof.sec}). 
In this case these observed features are due to two counter-rotating disks. Although this
particular scenario is unlikely, as
the line profile would be flat-topped rather
than peaked (\S\,\ref{h3h4.sec}), this illustrates the difficulty in interpreting
incomplete data in a unique way. 

\subsection{Radial velocity dispersion profile}
\label{sigprof.sec}

\begin{figure}[htbp]
  \begin{center}
  \includegraphics[width=0.95\linewidth]{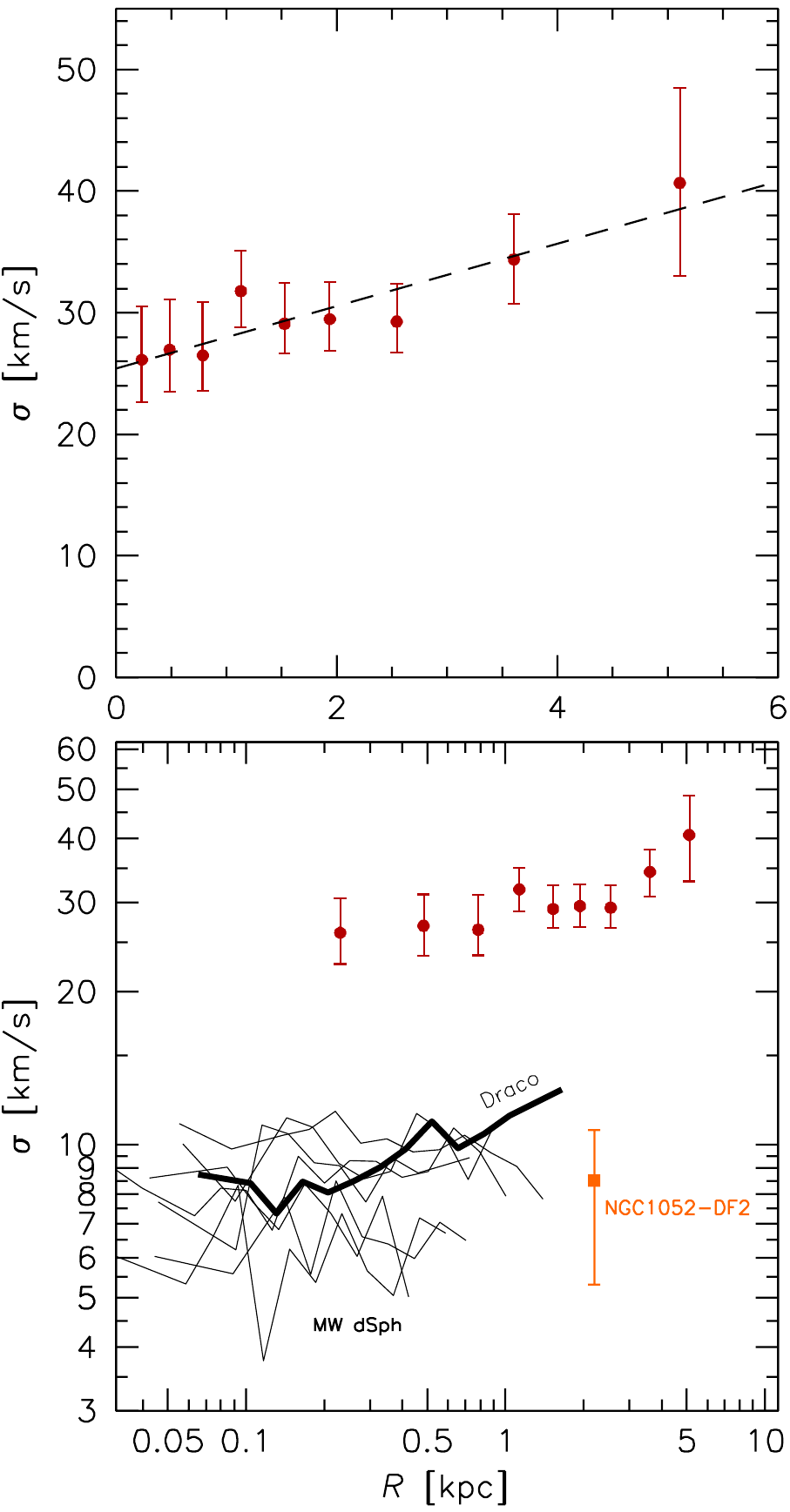}
  \end{center}
\vspace{-0.2cm}
    \caption{
Radial velocity dispersion profile of \gal, measured in
elliptical apertures. The radii are the luminosity-weighted
averages within each aperture.
The data in both panels are the same; the top panel is in linear
units and the bottom panel in logarithmic units. {\em Top panel:}
The dashed line is a
simple linear fit to the data. {\em Bottom panel:} The UDG NGC\,1052--DF2 is
shown by the orange point, from {Danieli} {et~al.} (2019).
Curves are data for seven classical Milky Way dwarf
spheroidals obtained from {Walker} {et~al.} (2007). Draco is highlighted as it has a
similarly
radially increasing profile as \gal.}
\label{ellipse.fig}
\end{figure}

Having established that the galaxy does not show appreciable rotation, we
assume that the kinematics can be meaningfully characterized in
the elliptical apertures shown in Figs.\ \ref{apertures.fig} and \ref{radial.fig}. 
The radial velocity dispersion profile is shown in Fig.\ \ref{ellipse.fig}, in linear
units (top panel) and logarithmic units (bottom panel). We find that the
velocity dispersion profile gradually increases with radius, and shows no
"second order" features such as a bump on small ($0.5-1$\,kpc) scales.
A simple linear fit gives
\begin{equation}
 \sigma= (24.9 \pm 1.0) + (2.9\pm 0.4) R,
\end{equation}
with $R$ in kpc and $\sigma$ in \kms. Note that only random errors are taken into account
in the uncertainties.

We include the UDG NGC\,1052--DF2 in Fig.\ \ref{ellipse.fig},
from {Danieli} {et~al.} (2019).
The new data for \gal\
confirm that the two galaxies 
have very different kinematics,
even though they have a similar luminosity,
morphology, and stellar population, and both have a relatively high number
of globular clusters. These two objects highlight the large object-to-object
scatter that appears to exist within the UDG population
(see also {Spekkens} \& {Karunakaran} 2018; {Toloba} {et~al.} 2018).

We also compare the \gal\ measurements to radial velocity
dispersion profiles
of Local Group dwarf spheroidal galaxies (dSphs). UDGs resemble dSphs
in terms of their visual
morphology, mean Sersic
index, axis ratio distribution, surface brightness, and high dark matter
fraction within the effective radius; the main (baryonic) differences
are their $\sim 10\times$ larger sizes, corresponding $\sim 100\times$ larger
luminosities and stellar masses, and their higher average globular cluster 
specific frequency (see, e.g., {van Dokkum} {et~al.} 2015; {Lim} {et~al.} 2018). 
Thin curves in the bottom panel of Fig.\ \ref{ellipse.fig} show the radial dispersion
profiles of seven classical dSphs,
obtained from kinematic data of
{Walker} {et~al.} (2007). The velocities and velocity dispersions
were added in quadrature, and rebinned
to logarithmic bins with a size of 0.1 dex. 
The radial profile of \gal\ is not very different from scaled-up versions of
those of
dSph galaxies. In particular, Draco has a positive dispersion gradient that
is similar to that of \gal.

\noindent
\begin{deluxetable}{ccrr}
\tablecaption{Velocity Dispersion Profile\tablenotemark{a}
\label{disp.table}}
\tablehead{\colhead{$R$\tablenotemark{b}} & \colhead{$R$}  & \colhead{$v$} & \colhead{$\sigma$}\\
\colhead{[arcsec]} & \colhead{[kpc]} & \colhead{[km\,s$^{-1}$]} & \colhead{[km\,s$^{-1}$]}}
\startdata
$0\farcs 5$ & 0.23 & $0.2_{-2.6}^{+2.2}$ & $26.1^{+4.4}_{-3.5}$  \\
$1\farcs 0$ & 0.49 & $-3.4_{-2.8}^{+2.6}$ & $26.7^{+4.1}_{-3.4}$ \\
$1\farcs 6$  & 0.79  & $-3.1_{-2.5}^{+2.2}$ & $26.5^{+4.4}_{-2.9}$ \\
$2\farcs 3$  & 1.13 & $-0.7_{-1.8}^{+2.0}$ & $31.8^{+3.3}_{-2.9}$ \\
$3\farcs 2$  & 1.53 & $0.5_{-2.6}^{+2.0}$ & $29.1^{+3.4}_{-2.4}$ \\
$4\farcs 0$  & 1.94 & $0.7_{-2.0}^{+2.0}$ & $29.5^{+3.0}_{-2.6}$ \\
$5\farcs 3$  & 2.55 & $-0.4_{-2.0}^{+2.3}$ & $29.3^{+3.1}_{-2.5}$ \\
$7\farcs 4$ & 3.62 &  $0.3_{-2.5}^{+2.3}$ & $34.4^{+3.8}_{-3.6}$ \\
$10\farcs 6$ & 5.13 & $5.9_{-4.1}^{+3.8}$ & $40.2^{+7.9}_{-7.6}$ 
\enddata
\tablenotetext{a}{The data in Fig.\ \ref{ellipse.fig} are $\sigma_{\rm eff}
= (\sigma^2+v^2)^{0.5}$.}
\tablenotetext{b}{Luminosity-weighted average radius of elliptical aperture; $R_{\rm maj}
\approx 1.20 R$.}
\end{deluxetable}

\begin{figure*}[htbp]
  \begin{center}
  \includegraphics[width=0.8\linewidth]{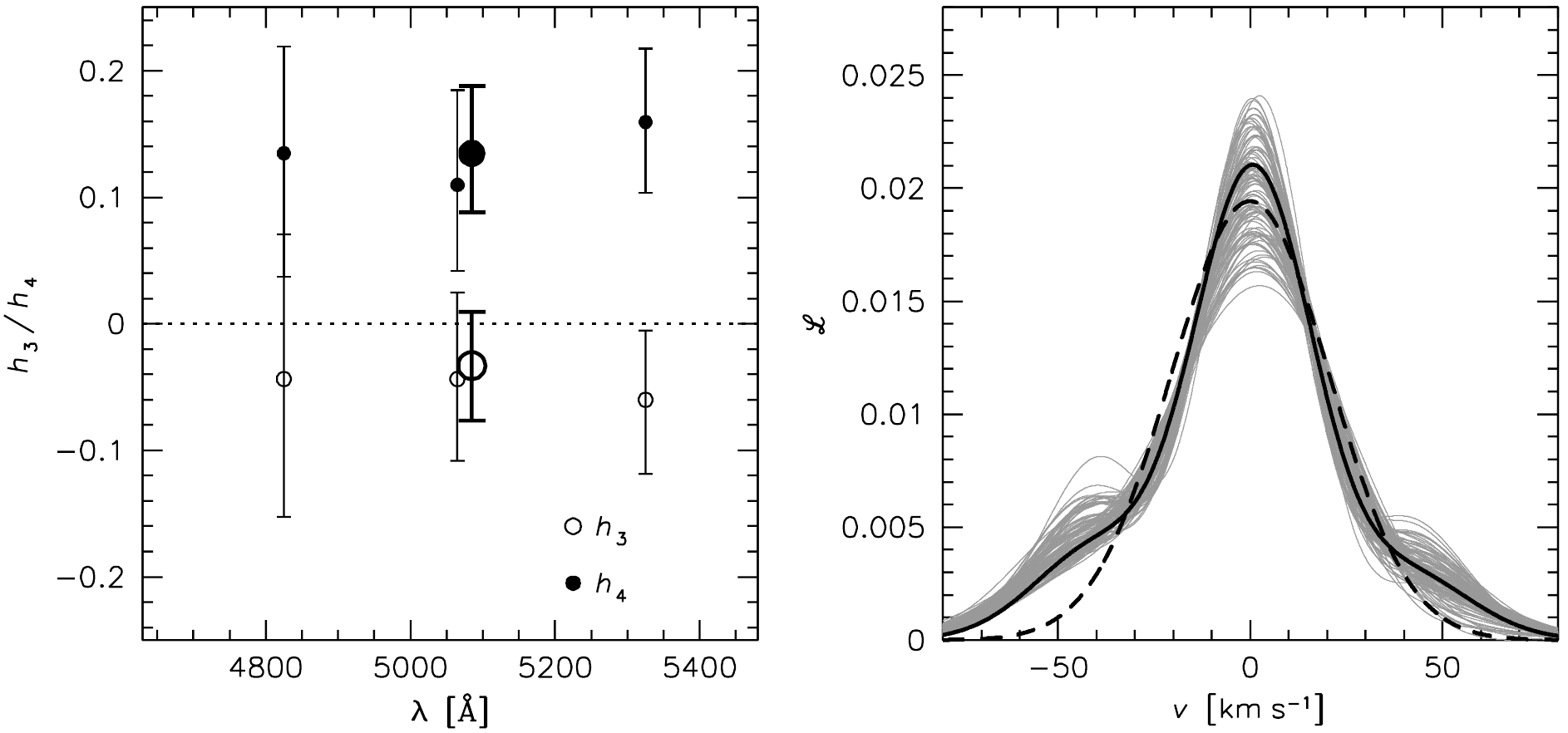}
  \end{center}
\vspace{-0.2cm}
    \caption{
Deviations from a Gaussian absorption line profile, as determined from the fit to the
optimally-extracted spectrum of \gal. The left panel shows the Gauss-Hermite $h_3$
and $h_4$ components, fitted in three wavelength intervals (small symbols) and for
the whole spectrum (large symbols). The best-fit values for the full wavelength
range are $h_3 = -0.03 \pm 0.04$ and $h_4 = 0.13 \pm 0.05$. 
The right panel shows the implied line profile, along with the best-fitting
Gaussian (dashed line), and 100 MCMC samples (light grey).
}
\label{lineprof.fig}
\end{figure*}

\subsection{Line profile}
\label{h3h4.sec}

The  line profile of \gal\ is non-Gaussian, with reasonably high significance. In Fig.\
\ref{lineprof.fig} the best-fitting $h_3$ and $h_4$ parameters are shown, for the
fit to the optimally-extracted spectrum.  We find $h_3=-0.03\pm 0.04$ and
$h_4 = 0.13 \pm 0.05$.
The $h_3$ parameter is consistent with zero,
indicating that the line profile is close to symmetric. This is necessarily
the case if the
galaxy is axisymmetric and
in dynamical equilibrium, as any asymmetric deviations on one side
of the galaxy should be inverted on the opposite side (see, e.g., Fig.\ 15 in {Bender} {et~al.} 1994).

The $h_4$ parameter, which measures symmetric deviations from a Gaussian, is
positive with a formal significance of $3\sigma$. We performed many tests to assess
whether this result is reliable or driven by some subtle systematic error in the
analysis. The result persists when the spectrum is split in separate wavelength
intervals (see Fig.\ \ref{lineprof.fig}) or when $h_3$ is forced to be zero;
it is seen in almost all radial bins
(albeit with low significance for each individual bin); it persists when the
instrumental line profile is assumed to be Gaussian instead of flat-topped; it
is not sensitive to the wavelength calibration corrections described in \S\,\ref{combinespecs.sec}
and \S\,\ref{fitting.sec}; and it is insensitive to the exact template that is used.

A positive $h_4$ can indicate radial orbital anisotropy
(see, e.g., {Bender} {et~al.} 1994; {Thomas} {et~al.} 2007; {Amorisco} \& {Evans} 2012) but can also be due
to other effects, such as
flatteningf of the galaxy along the line of sight (e.g., {Magorrian} \& {Ballantyne} 2001). 
Although there is no unique interpretation of the $h_4$ measurement it is
difficult
to reconcile with {\em tangential} anisotropy, which is required to fit the radial velocity
dispersion profile for certain halo models (as we show later).

\subsection{Robustness of the radial dispersion profile}

We end this section with an empirical assessment of the accuracy of the error bars on the
dispersion measurements. The residuals from the template fits to the spectra are consistent 
with the errors in the data, but as discussed
in \S\,4.3 this does not address possible systematic errors in the dispersion
measurements. Furthermore, the interpretation of the uncertainties is complicated by the fact that
the spectra were resampled to a finer grid.

The major and minor axis kinematics of \S\,5.1 provide a test that relies only on the assumption that
the dispersion profiles are symmetric with respect to the center of the galaxy.
The major axis profile  includes seven dispersion measurements on the ``left'' side and
seven measurements on the ``right''. Similarly, the minor axis profile includes six measurements on either side of the galaxy's center.
Assuming that the profile is symmetric, we therefore have thirteen pairs of independent measurements of the same quantity, determined from regions that span the entire detector.
The scatter in the differences between these paired dispersions is $5.7\pm 1.1$\,\kms.
The expected variation is $\langle (e_{1,i}^2 + e_{2,i}^2)^{0.5} \rangle_i = 6.2\pm 0.4$\,\kms, where
$e_1$ and $e_2$ are the uncertainties in the measurements on the two sides of the galaxy.
The observed variation is consistent with the expected variation, and we conclude that the quoted
uncertainties accurately reflect the observed variation between independent
measurements.

\section{Dynamical Mass and M/L Ratio}
\label{mass.sec}

\subsection{Is \gal\ in dynamical and structural equilibrium?}
\label{environment.sec}

Before interpreting the kinematics
we first ask whether the
galaxy is in equilibrium. Some UDGs clearly are not, with the 
``boomerang galaxy'' M101-DF4 ({Merritt} {et~al.} 2016) a case in point, and
it has been suggested
that many Coma UDGs are in the process of tidal disruption by the
cluster potential
({Yozin} \& {Bekki} 2015). The rising velocity dispersion profile of
\gal\ could be interpreted as evidence for such disruption,
as unbound material
may inflate the observed velocity dispersion at large radii. A demonstration
of this was given by {Mu{\~n}oz}, {Majewski}, \&  {Johnston} (2008), who reproduced the rising
velocity dispersion profile of the Carina dSph in
the context of such models.

However, 
rising velocity dispersion profiles such as that of Carina,
Draco and \gal\ do not {\em necessarily} imply tidal disruption.
In fact, the kinematics of
Draco have been reproduced with relatively simple
mass models ({Kleyna} {et~al.} 2002), and the galaxy has
been described as ``flawless'' because
of its lack of detected tidal features in deep imaging data ({S{\'e}gall} {et~al.} 2007).
Furthermore, as noted in \S\,1, tidal disruption
scenarios cannot be easily reconciled with
the high globular cluster counts of UDGs 
in Coma ({Lim} {et~al.} 2018) and their lack of obvious
tidal features ({Mowla} {et~al.} 2017).

In the specific case of
\gal\ tidal heating by the Coma
cluster is unlikely because it  appears to live in a dynamically-cold
environment. Three other low surface brightness galaxies in the
vicinity of \gal\ (Dragonfly~42, DFX1, and DFX2) have redshifts,
all from the DEIMOS multi-slit
spectroscopy described in {van Dokkum} {et~al.} (2016, 2017a),
and {Alabi} {et~al.} (2018). Of this sample of four faint galaxies,
three
(Dragonfly~42, Dragonfly~44, and DFX2) are within $\approx 100$\,\kms\ of each
other
(see {van Dokkum} {et~al.} 2017a; {Alabi} {et~al.} 2018).
The velocity dispersion of the Coma cluster
is $\sim 1000$\,\kms\ ({Colless} \& {Dunn} 1996), and the probability that three
out
of four randomly selected galaxies have a velocity range $\Delta v<100$\,\kms\
is $3\times 10^{-3}$.

It is unclear whether \gal\ is in a cold clump that is falling into
the cluster, a filament, or a structure that is unrelated to Coma;
this can be constrained by measureing redshifts for more galaxies in the
vicinity of \gal. Irrespective of the precise interpretation,
the small redshift range strongly suggests that the galaxy
has not been part of the Coma cluster for a long time, and is therefore unlikely
to be significantly
affected by tidal heating or other cluster-driven
processes.  Although it is difficult to completely rule out tidal effects,
in the following
we assume that the galaxy is in equilibrium and
that its dynamics reflect the galaxy's gravitational potential.

\begin{figure*}[htbp]
  \begin{center}
  \includegraphics[width=0.9\linewidth]{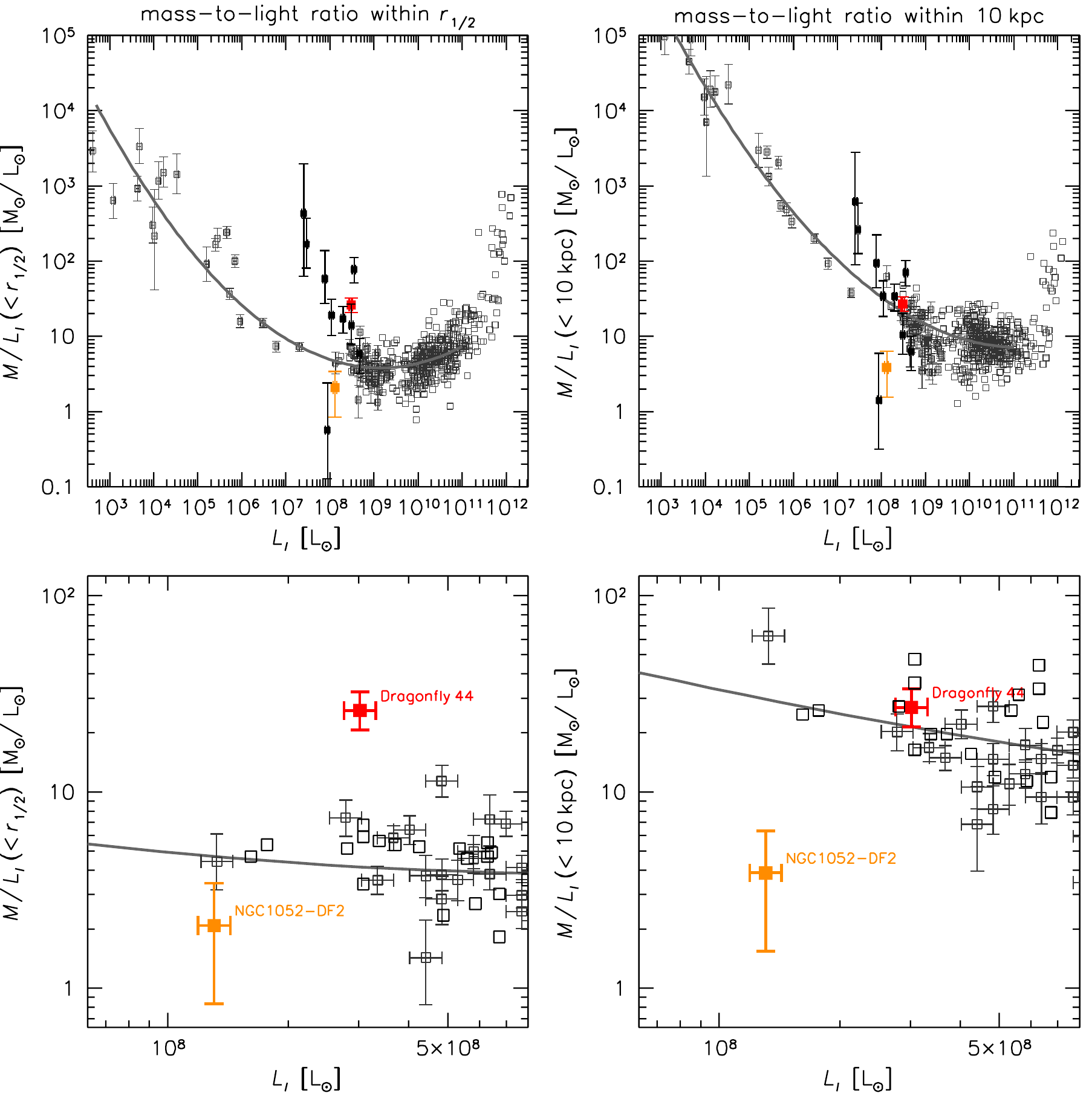}
  \end{center}
\vspace{-0.2cm}
    \caption{
Relations between dynamical $M/L_I$ ratio and total luminosity, $L_I$.
Grey points are normal galaxies; black and colored points
are UDGs, with \gal\ indicated with the red symbol.
Literature sources are given in the text. The bottom panels show a small
section of the top panels, focusing on \gal\ and NGC\,1052-DF2.
{\em Left panels:} $M/L_I$ ratio within the half-light radius.
\gal\ has a high $M/L$ ratio; similar to other
UDGs it is dark matter-dominated
within its half-light radius. {\rm Right panels:} $M/L_I$ radius
within a fixed radius of 10 kpc (see text). Now the UDGs, including
\gal, fall within
the distribution of other galaxies. We infer that the high
$M/L$ ratio of \gal\  within $R_{\rm e}$ mostly reflects its large effective
radius, not necessarily an unusually high dark matter mass on kpc scales.
}
\label{wolf.fig}
\end{figure*}

\subsection{Dynamical mass within the effective radius}
\label{mlre.sec}

The velocity dispersion profile extends slightly beyond the projected
half-light radius of \gal, $R_{\rm e, maj}=4.7$\,kpc. As shown in 
{Wolf} {et~al.} (2010), the luminosity-weighted velocity dispersion within
the projected circularized
half-light radius $R_{\rm e,c}=R_{\rm e, maj}(b/a)^{0.5}$
provides a robust estimate of the dynamical
mass within the 3D half-light radius $r_{1/2}$ that is insensitive to
anisotropy or the form of the density profile.
We measure the luminosity-weighted dispersion directly from a
luminosity-weighted extracted spectrum within the half-light radius.
The resulting dispersion is $\sigma_{\rm e} = 33^{+3}_{-3}$\,\kms. This value is
lower than that reported in {van Dokkum} {et~al.} (2016), who found
$\sigma=47_{-6}^{+8}$\,\kms\ based on a DEIMOS spectrum in the H$\alpha$
region. A re-assessment of the 2016 analysis uncovered an error;
the revised DEIMOS dispersion is $\sigma=42^{+7}_{-7}$\,\kms, closer to
the KCWI value.\footnote{In the 2016 analysis the spectroscopic data were combined
without applying barycentric velocity corrections to each individual dataset.
However, as the DEIMOS data were taken over a period of several
months the peak-to-peak velocity corrections are $\approx 30$\,\kms.
As a result, the combined spectrum was slightly broadened, leading to
a dispersion measurement that was biased high. After applying the required
corrections we derive $\sigma=42$\,\kms\ instead of 47\,\kms.}
Despite this better agreement, the probability that the difference
can be attributed to chance is only 3.3\,\%. We do not have
an explanation for the discrepancy but speculate that
it may be caused by the large weight of the Balmer
H$\alpha$ line in the {van Dokkum} {et~al.} (2016) analysis or systematic
errors introduced by the crosstalk corrections that were needed.

The {Wolf} {et~al.} (2010) estimator,
\begin{equation}
M(r<r_{1/2}) \approx 9.3\times 10^5\sigma_{\rm e}^2 R_{\rm e,c},
\end{equation}
gives $M(r<r_{1/2}) = 3.9_{-0.5}^{+0.5} \times 10^9$\,\msun.
The total $I_{814}$ magnitude of \gal\ is $M_I = -16.7$
({van Dokkum} {et~al.} 2017a), or $L_I = 3.0^{+0.6}_{-0.6} \times 10^8$\,L$_{\odot}$,
assuming a 20\,\% error in the total luminosity.
Therefore, the $M/L_I$ ratio within the 3D half light radius is
$M/L_I(r<r_{1/2}) = 26^{+7}_{-6}$\,M$_{\odot}$/L$_{\odot}$.

The expected $M/L_I$ ratio from the stellar population is
$M/L_I \approx 1-1.5$, and the implication is that \gal\ is extremely
dark matter dominated within its half radius. This is generally
not the case for galaxies in this mass and luminosity range,
as demonstrated in the left panels of Fig.\ \ref{wolf.fig}.
Here we show the relation between the dynamical $M/L_I$ ratio
within the half-light radius as a function of the total luminosity.
Grey points show samples
of ``normal'' galaxies from {Zaritsky}, {Gonzalez}, \&  {Zabludoff} (2006), {Wolf} {et~al.} (2010), and
{Toloba} {et~al.} (2015). The line is a fit to these samples for
$L_I<10^{11}$\,L$_{\odot}$, of the form
\begin{equation}
\log M/L_I(<r_{1/2})= 0.69 - 0.194 L_8 + 0.0838 (L_8)^2,
\end{equation}
with $L_8\equiv \log L_I - 8$, $M/L_I(<r_{1/2})$ the dynamical mass-to-light ratio within
the half-light radius, and $L_I$ the total $I$ band luminosity. The rms scatter
around this line is only 0.22\,dex, independent of luminosity.

Large black and colored
points are UDGs from {Beasley} {et~al.} (2016), {Toloba} {et~al.} (2018),
{Mart{\'{\i}}n-Navarro} {et~al.} (2019), {Danieli} {et~al.} (2019), {van Dokkum} {et~al.} (2019),
and {Chilingarian} {et~al.} (2019).\footnote{We only show objects with $R_{\rm e}>1.5$\,kpc
from {Chilingarian} {et~al.} (2019).} This is an
update to Fig.\ 3 in {van Dokkum} {et~al.} (2016) and Fig.\ 4 in
{Toloba} {et~al.} (2018), which showed the relation
between $M/L$ ratio and dynamical mass. \gal, and other
UDGs with measured kinematics, are in a
``no man's land'' in this parameter space, with $M/L$ ratios
that are similar to much fainter and much brighter galaxies.
Phrased differently,
the small scatter in the well-known U-shaped distribution of
galaxies in this plane is likely partially due to selection effects.

\subsection{Dynamical mass within a fixed radius}
\label{ml10.sec}

It is tempting to interpret
the vertical axis of the left panels of
Fig.\ \ref{wolf.fig} in terms of the ratio
between total halo mass and total stellar mass. This is often done
implicitly; e.g., {Martin} {et~al.} (2018) use the $M/L$ ratio within
the effective radius as a probe of the total halo mass when assessing
whether the UDG NGC\,1052-DF2 is lacking in dark matter. 
However, a second parameter, the effective radius,
plays an important role. {\em The effective radius
always contains 50\,\% of the light, but it does not contain
a fixed fraction of the dark matter.} At fixed halo mass, virial
radius, and concentration, the enclosed dark matter mass within the half-light radius
(and therefore the dynamical $M/L$ ratio) is expected
to scale with that radius.\footnote{The same is true if the mass were measured within, say,
$2R_{\rm e}$ or some other measure of the optical extent of the galaxies.}

We assess this effect by estimating the $M/L_I$ ratio within
a fixed 3D radius of $r=10$\,kpc for all the galaxies in the samples
quoted above.
The mass is extrapolated by assuming a flat rotation curve, that is,
\begin{equation}
M(r<10\,{\rm kpc}) = \frac{10}{r_{1/2}} M(r<r_{1/2}),
\end{equation}
with $r_{1/2} \approx 4/3 R_{\rm e,c}$.
The luminosity is extrapolated by numerically
integrating the {Sersic} (1968)
profile out to $r=10$\,kpc, where the Sersic index is assumed to be

\begin{equation}
n = 
\begin{cases}
  1 & \text{if $L_I<10^{10}$\,L$_{\odot}$} \\
  1+2.5\left[\log(L_I)-10\right] & \text{otherwise}
\end{cases}.
\end{equation}

\begin{figure*}[htbp]
  \begin{center}
  \includegraphics[width=0.85\linewidth]{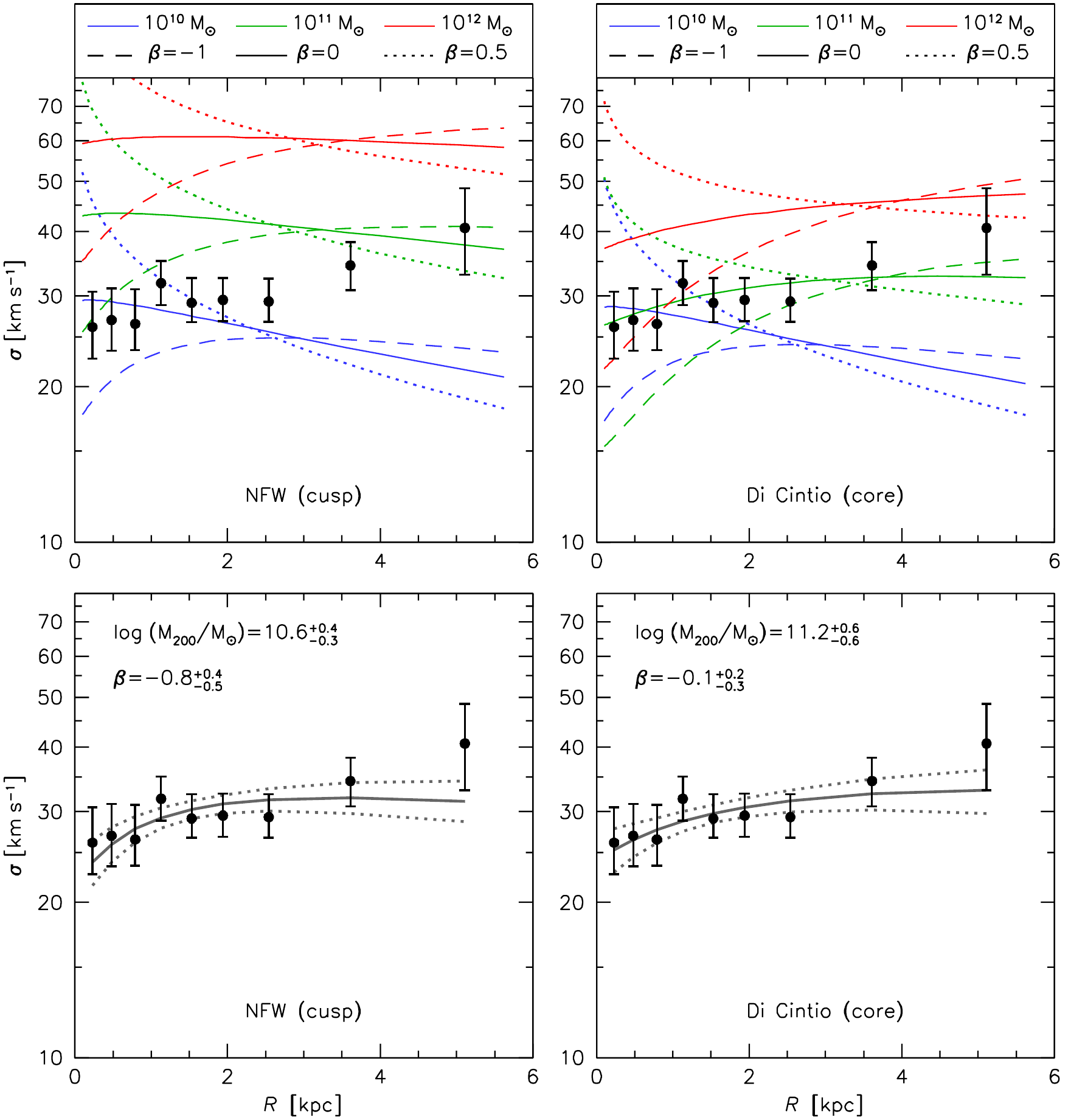}
  \end{center}
\vspace{-0.2cm}
    \caption{
Dark matter halo fits to the velocity dispersion profile. {\em Top panels:} 
Illustration of velocity dispersion profiles for various model assumptions:
standard NFW halos (left) versus profiles with a mass-dependent
core from {Di Cintio} {et~al.} (2014a) (right); different halo masses
(colors); and radial ($\beta=0.5$) versus tangential ($\beta=-1$) anisotropy.
The curves are not fits to the data but
do take the observed surface brightness profile of the galaxy into account.
{\em Bottom panels:} Best fits to the observed kinematics. The halo mass
and anisotropy are dependendent on the assumed density profile of the halo.
The halo mass is lower in cuspy
models than in cored models.
NFW halos require strong tangential anisotropy to explain the rising velocity
dispersion
profile, whereas Di Cintio halos do not. 
}
\label{models_data.fig}
\end{figure*}

The results are shown in the right panels of Fig.\ \ref{wolf.fig}.
Low luminosity galaxies tend to have small effective radii, and their
$M/L$ ratios within 10\,kpc are much higher than those within $r_{1/2}$.
High luminosity galaxies have $r_{1/2}\sim 10$\,kpc, and their
$M/L$ ratios within 10\,kpc are similar to
those within $r_{1/2}$. Because of this correlation of the
effective radius with luminosity,
the distribution of 
galaxies in the right panels
is very different than in the left panel.
The best fitting relation,
\begin{equation}
\log M/L_I(<10\,{\rm kpc})
= 1.52 - 0.427 L_8 + 0.0682 (L_8)^2,
\end{equation}
has a factor of seven higher normalization at $L=10^8$\,L$_{\odot}$ than the relation
between $M/L(<r_{1/2})$ and luminosity. 

However, \gal, as well as other UDGs, stay at nearly the same location.
As a result, \gal\ is now consistent with the relation defined by normal
galaxies,
whereas NGC\,1052-DF2 now falls far below it: in the left panels
its $M/L$ ratio is similar to that of other galaxies, but given
its large effective radius its $M/L$ ratio should have been
much higher if it had a normal dark matter halo.
We conclude that it is hazardous to interpret the $M/L$ ratio within the
effective radius in terms of halo masses. The $M/L$ ratio within a fixed large
aperture should provide a better indication, but for most galaxies in
Fig.\ \ref{wolf.fig} this represents a significant extrapolation beyond the
regime where the kinematics are measured.


\begin{figure*}[htbp]
  \begin{center}
  \includegraphics[width=0.85\linewidth]{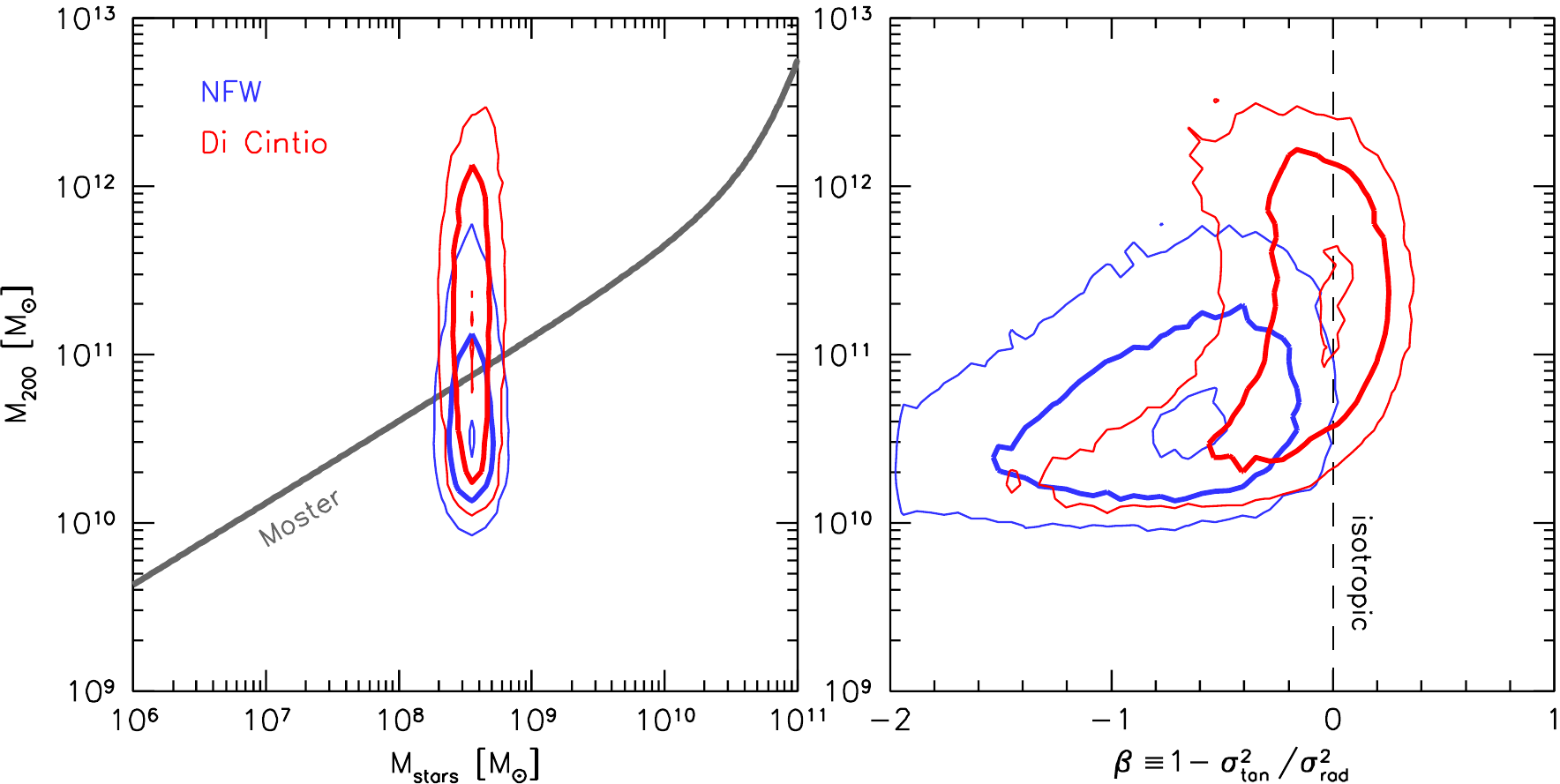}
  \end{center}
\vspace{-0.2cm}
    \caption{
Joint constraints on the halo mass and stellar mass (left) and the halo mass
and the anisotropy parameter $\beta$ (right), for the two different halo
profiles that are considered here. The thick line encloses 68\,\% of the
samples, and the outer contour encloses 95\,\% of the samples. In the left panel the
stellar mass -- halo mass relation of {Moster} {et~al.} (2010) is shown for reference.
}
\label{constraints.fig}
\end{figure*}

\section{Dark matter halo fits}
\label{halofits.sec}

Here we seek to interpret the measured kinematics in the context of
parameterized models for the mass distribution. In particular, we ask what classes
of models can reproduce the rising velocity dispersion profile and the
positive $h_4$ parameter, and what the implications are for the halo mass
of \gal.

\subsection{Procedure}

We use
the methodology that is outlined in {Wasserman} {et~al.} (2018b)
and {Wasserman} {et~al.} (2018a).
Briefly, spherical mass models with a given density profile are fit to the
observed velocity dispersion profile using a Bayesian Jeans modeling formalism.
The mass distribution is modeled as the sum of the stellar distribution
and a parameterized dark matter halo profile.
For the halo we use two descriptions that are both
instances of the general $(\alpha,\beta,\gamma)$ profile,
\begin{equation}
\label{nfw.eq}
\rho(r) = \rho_s \left(\frac{r}{r_s}\right)^{-\gamma}
 \left[ 1 + \left( \frac{r}{r_s}\right)^{\alpha} \right]^{\frac{\gamma-\beta}{\alpha}},
\end{equation}
with $r_s$ the scale radius and $\rho_s$ the scale density
({Hernquist} 1990). These profiles have a powerlaw slope $-\gamma$ on small
scales and $-\beta$ on large scales, with the form of the profile near
the transition controlled by $\alpha$.

The first model is a standard
cuspy  Navarro-Frenk-White (NFW) profile, with $\alpha=1$, $\gamma=1$, and $\beta=3$
({Navarro} {et~al.} 1997).
The second is a cored model with a flatter inner density profile.
This can be achieved by reducing the value of $\gamma$ in Eq.\ \ref{nfw.eq}
while retaining $\alpha=1$ and $\beta=3$
(e.g., {Zhao} 1996; {Wyithe}, {Turner}, \& {Spergel} 2001), or by using physically-motivated fitting functions
such as the ``CORENFW" profile of {Read}, {Agertz}, \& {Collins} (2016). 
Here we use the parameterization of {Di Cintio} {et~al.} (2014a), who use high
resolution hydrodynamical simulations to derive ``empirical'' relations between
the stellar-to-halo mass $X = \log( M_{\rm stars}/M_{\rm halo})$ and $(\alpha,\beta,\gamma)$.
These relations  are specified in Eq.\ 3 of
{Di Cintio} {et~al.} (2014a). The relation for the inner density profile is
given by
\begin{equation}
\gamma = -0.06 + \log\left[\left(10^{X+2.56}\right)^{-0.68} + 10^{X+2.56}\right].
\end{equation}
The Di Cintio profile is similar to NFW for very low mass galaxies and for $L_*$ galaxies,
and has a core that is
maximal for stellar masses of $\sim 10^{8.5}$\,\msun, that is, the stellar mass
of \gal\ and of most other UDGs that have been studied in detail so far.
For this mass $X\sim -2.5$ (for UDGs that lie on the stellar mass -- halo mass relation)
and $\gamma \sim 0.3$.
We note that, as the stellar mass of \gal\
is fairly well constrained, this model does not have significantly more freedom than
the standard NFW profile.

Halo masses are expressed in terms of $M_{200}$ and the concentration
$c_{200}$, with
\begin{equation}
M_{200} = 200\rho_{\rm crit} \frac{4\pi r_{200}^3}{3},
\end{equation}
with $c_{200}=r_{200}/r_s$.
We assume the median concentration
from the halo mass -- concentration relation
determined by {Diemer} \& {Kravtsov} (2015).
In addition
to the halo parameters the anisotropy
$\beta = 1- \sigma_{\rm tan}^2/\sigma_{\rm rad}^2$ is a fit parameter.\footnote{We
use two different parameters in this paper that are both denoted ``$\beta$'': the
second coefficient in the ($\alpha$, $\beta$, $\gamma$) profile (Eq.\ 21), and
Binney's anisotropy parameter. These are both conventional expressions, and we believe
changing either of them would be confusion. Hopefully it is always clear from context which $\beta$ the text is referring to.} For
simplicity the anisotropy is assumed to be constant with radius; models with varying anisotropy
give qualitatively similar results, albeit with more freedom in the mass profiles.

The top panels of Fig.\ \ref{models_data.fig} illustrate the behavior of
the models and what parameters can be constrained by the data. The curves
are based on Jeans modeling with fixed model parameters; that is, they
are not fits to the data but they do take the observed surface brightness
profile of \gal\ into account.
More massive halos obviously produce higher velocity dispersions, but the
effect is relatively small: about a factor of $\approx 1.5$
change in velocity dispersion for a factor of 10 change in halo mass.
At fixed halo mass and anisotropy
the predicted dispersions are higher
for NFW halos than for the cored Di Cintio halos, although the
difference vanishes for low halo masses.
The Di Cintio halos readily predict rising velocity dispersion profiles,
particularly for halo masses $M_{200}\sim 10^{11}$\,\msun\ where the cores are
maximal.
However, the
shape of the velocity dispersion profile is degenerate with
the anisotropy parameter $\beta$:
generically, radial anisotropy produces falling
profiles whereas tangential anisotropy produces rising profiles. 
In principle, this degeneracy can be resolved
by including the form of the absorption line profile
in the  
analysis
(see \S\,5.3 and also, e.g., {Amorisco} \& {Evans} 2012).
We will return to this below.

\subsection{Results}

The models are fit to the data using the {\tt emcee} MCMC sampler
({Foreman-Mackey} {et~al.} 2013), as
described in {Wasserman} {et~al.} (2018b).
The following priors are used:
\begin{eqnarray}
P(\log (M_{\rm stars}/L_V)) & = & N(\log(1.5), 0.1^2)
\label{mlprior.eq}\\
P(\log (M_{200})) & = & U(7,15)\\
P(-\log(1-\beta)) & = & U(-1.5,1.5),
\end{eqnarray}
where $N(\mu,\sigma^2)$ is the normal distribution and $U({\rm min},{\rm max})$ is
a uniform distribution. The only informative prior is Eq.\ \ref{mlprior.eq}.
The mean $M/L_V$ ratio comes from stellar population
synthesis modeling (e.g., {van Dokkum} {et~al.} 2016), with a standard deviation
obtained from {Taylor} {et~al.} (2011).

Good fits are obtained for both standard NFW halos and Di Cintio cored halos,
as shown in the bottom panels of Fig.\ \ref{models_data.fig}. The distributions
of MCMC samples for $M_{200}$ and $\beta$ are shown
in Fig.\ \ref{constraints.fig}. 
NFW halos
require strong tangential anisotropy whereas the
Di Cintio profiles do not. For an NFW profile
the best fitting halo mass is $\log (M_{200}/{\rm M}_{\odot}) = 10.6^{+0.4}_{-0.3}$
and $\beta=-0.8^{+0.4}_{-0.5}$, whereas these values are
$11.2^{+0.6}_{-0.6}$ and $-0.1^{+0.2}_{-0.3}$ respectively for the Di Cintio
profile. One way to view these results is that the Di Cintio models ``naturally'' 
predict rising velocity
dispersion profiles for the stellar mass regime of \gal, whereas
NFW profiles predict decreasing profiles unless strong tangential anisotropy is
invoked.

As shown in Fig.\ \ref{constraints.fig} \gal\
is consistent with the stellar mass -- halo mass relation of {Moster} {et~al.} (2010)
within $1\sigma$;
this relation is very similar to that of {Behroozi} {et~al.} (2013a) and others in this
regime.
However, 
the total halo mass is not particularly well constrained 
in either model, as the data
sample the halo only out to 
a small fraction of the virial radius. The uncertainty in the halo mass is
much larger than the probable scatter of 0.2\,dex in the relation (see,
e.g., {Behroozi}, {Wechsler}, \&  {Conroy} 2013b; {Gu}, {Conroy}, \& {Behroozi} 2016). The halo mass is higher for a
Di Cintio profile than for an NFW profile;
in particular, a halo mass of $10^{12}$\,\msun, as is indicated by the globular cluster
counts of \gal\ ({van Dokkum} {et~al.} 2016, 2017a; {Harris}, {Blakeslee}, \&  {Harris} 2017; {Forbes} {et~al.} 2018), is within 
the $1\sigma$ contour for the Di Cintio
model whereas it
falls outside of the $2\sigma$ contour for the NFW model.
We note that the {\em stellar} mass is not constrained by the kinematics;
the distribution of the MCMC samples simply follows the prior.

\begin{figure}[htbp]
  \begin{center}
  \includegraphics[width=0.92\linewidth]{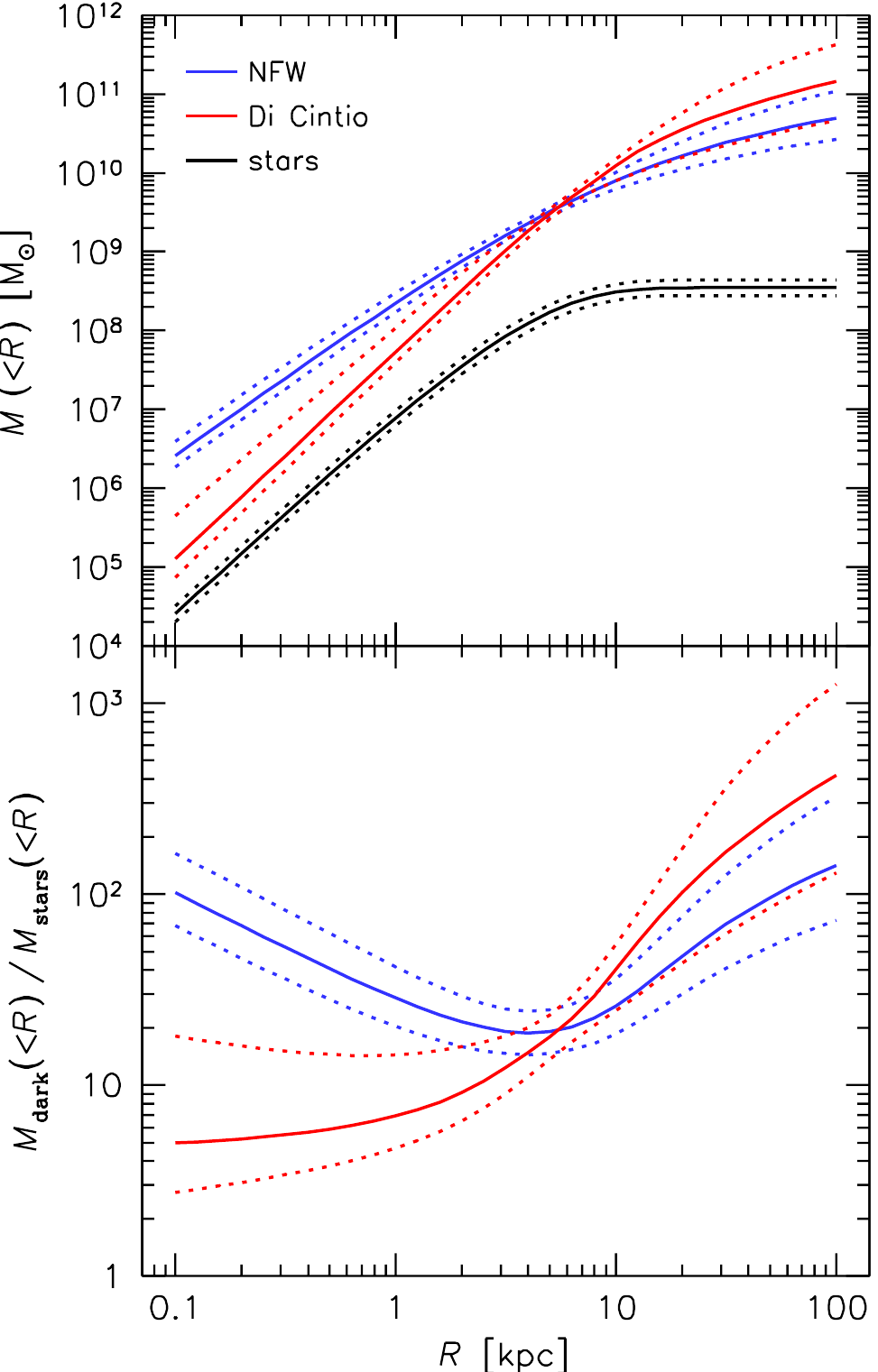}
  \end{center}
\vspace{-0.2cm}
    \caption{
{\em Top panel:}
Enclosed mass in dark matter and stars
as a function of radius in \gal, for NFW
halos  and cored halos.
{\em Bottom panel:}
Ratio of enclosed dark matter mass and stellar mass.
Due to the low
Sersic index and large effective radius of the galaxy the dark
matter fraction
is high at all radii, even for cored halos.
This makes it possible to study the inner dark matter
profile in a relatively unambiguous way, in a halo mass regime where galaxies
typically have a significant contribution from baryons in their centers.
}
\label{fraction.fig}
\end{figure}

The integrated mass profiles for both types of halos are shown in Fig.\
\ref{fraction.fig}, along with the stellar mass profile.
The galaxy is dominated by dark matter at all radii even in the cored Di Cintio
model. The dark matter
fraction within 1 kpc is somewhat model dependent, but it is
well-constrained at $f_{\rm dm}\sim 95$\,\% on scales of a few kpc where we have
the most information.
These results extend the analysis of \S\,\ref{mass.sec}, where we showed that
\gal\ has the expected $M/L$ ratio within its effective radius for a ``normal''
dark matter halo. UDGs {\em should} have very high $M/L$ ratios, because they
are so large:  galaxies such
as NGC\,1052-DF2, with a $M/L$ ratio within the effective radius
that is not very different from other galaxies
of the same luminosity (Fig.\ 13, and {Danieli} {et~al.} 2019),
are the outliers.

\subsection{Anisotropy or core?}

We have shown that the velocity dispersion profile of \gal\ can be reproduced with two
classes of models: a
standard cuspy
NFW profile combined with strong tangential anisotropy,  or a cored
profile that is close to isotropic. 
As discussed in \S\,\ref{h3h4.sec} the shape of the absorption line profile can,
in principle, constrain the degree of anisotropy and therefore help decide which of
these two options is more likely.
Generically, a positive $h_4$ parameter indicates {\em radial} rather
than tangential anisotropy, as
radial orbits 
create excesses at both zero velocity and in the wings of the velocity distribution
(see, e.g., Fig.\ 2 in {van der Marel} \& {Franx} 1993).
Qualitatively, the positive $h_4=0.13 \pm 0.05$ is inconsistent with both models, as neither
model has significant radial anisotropy (see the right
panel of Fig.\ \ref{constraints.fig}). It is {\em more inconsistent} with the NFW profile,
as this model requires strong tangential anisotropy.

\begin{figure}[htbp]
  \begin{center}
  \includegraphics[width=0.92\linewidth]{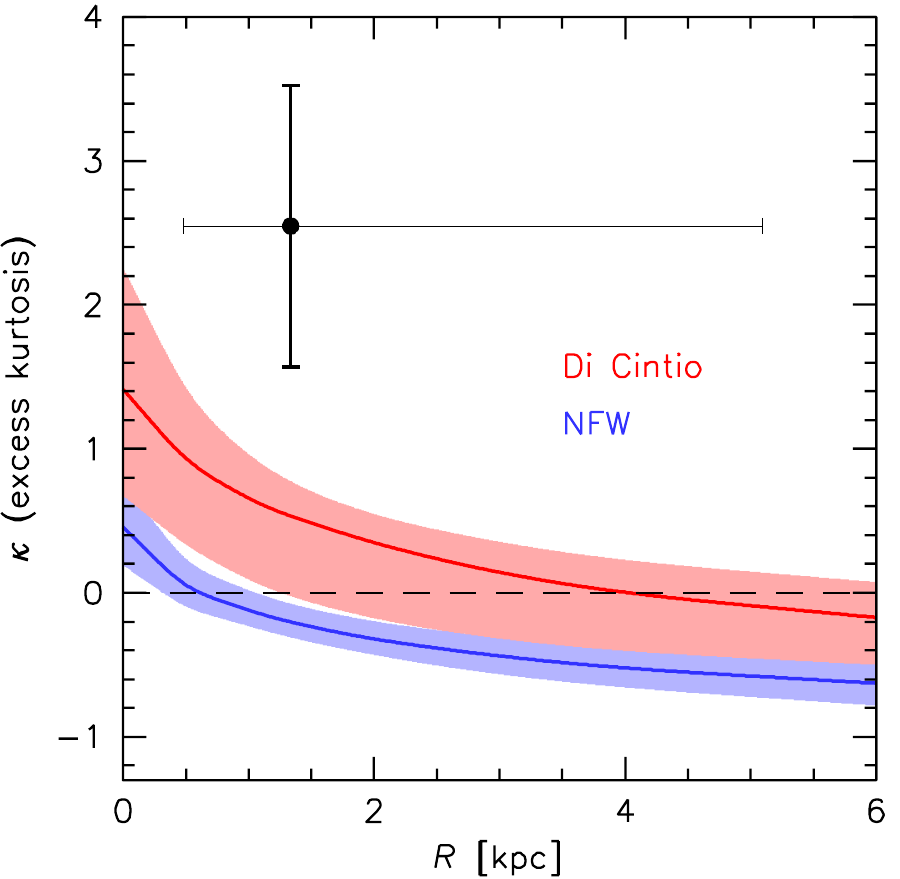}
  \end{center}
\vspace{-0.2cm}
    \caption{
Predicted symmetric deviations from a Gaussian line profile from our Jeans modeling
of \gal. The excess kurtosis $\kappa$ depends on radius, but is always
larger in the cored, approximately
isotropic, Di Cintio model than in the tangentially anisotropic NFW model.
The data point is the $h_4$ measurement from the optimally-extracted spectrum,
with $\kappa = 8\sqrt{6}h_4$. It is inconsistent with both models, although the distance
to Di Cintio is smaller than to NFW
($2.0\sigma$ versus $2.4\sigma$).
}
\label{kurtosis.fig}
\end{figure}

However, non-zero kurtosis can have other causes than anisotropy in the orbital
distribution, such as deviations from
spherical symmetry (e.g., {Read} \& {Steger} 2017) and the presence of a core in the
density profile. As shown by {{\L}okas} (2002), a cored profile leads to
positive kurtosis, and this may be the reason for the positive $h_4$
parameter that we measure.
We quantify this by
determining the kurtosis ($\xi_2 = \mu_4/\mu_2^2$, where $\mu_i$ is
the $i^{\rm th}$ moment)
and the excess kurtosis ($\kappa = \xi_2 - 3$) from the posteriors
of the model fits.
Figure\ \ref{kurtosis.fig} shows the radial dependence of $\kappa$ for both choices
of the density profile. The NFW model produces slightly negative kurtosis, as expected
from the tangential anisotropy. The cored model produces positive kurtosis despite its nearly isotropic orbital distribution,
as was also found by {{\L}okas} (2002).

The data point in Fig.\ \ref{kurtosis.fig} is the $h_4$ measurement from
the optimally-extracted spectrum. The weighted radius of this extraction
is 1.3\,kpc (as determined from the mean flux and weight of each spectrum
that contributes to it). The Gauss-Hermite coefficient
$h_4$ was converted to excess kurtosis
using $\kappa \approx 8\sqrt{6} h_4$ ({van der Marel} \& {Franx} 1993). The observed
kurtosis is higher than in either of the models, but closer to the
cored model than to the NFW one: the distance to the Di Cintio model
is $2.0\sigma$ and the distance to the NFW model is $2.4\sigma$.
This discrepancy may indicate that the central density profile is
even flatter than $\gamma \sim 0.3$, which it is in the Di Cintio
model. It could also reflect the limitations of the assumption of
spherical symmetry in the Jeans modeling (which is known to be
incorrect, as Dragonfly~44 has $b/a=0.69$; see also {Burkert} 2017), or even the assumption that the
galaxy is in equilibrium. Finally, we cannot exclude
undiagnosed systematic errors in the line profile
measurement.  With these caveats, we cautiously conclude that
the observed line profile is more consistent with a cored profile than
with an NFW profile.

\section{Discussion and Conclusions}

In this paper we present spatially-resolved kinematics of the UDG
\gal, obtained with
KCWI on the Keck\,II telescope. We find no evidence for rotation, which is
significant as \gal\ is one of the more flattened UDGs: its axis ratio
is $b/a=0.69$, whereas the median
for Coma UDGs is 0.74 ({van Dokkum} {et~al.} 2015). The limit that
we derive is more stringent than for many other low luminosity galaxies
(see Fig.\ \ref{vsig.fig}). This result is difficult to reconcile with
models in which UDGs are the high spin tail of the distribution of
normal dwarf galaxies, as was proposed by {Amorisco} \& {Loeb} (2016). Amorisco
et al.\ suggest processing by the cluster environment may decrease
$V/\sigma$, but as we discuss in \S\,\ref{environment.sec}
\gal\ appears to be in a dynamically-cold environment.

The velocity dispersion within the effective radius is lower than what
we reported in {van Dokkum} {et~al.} (2016), and
as discussed in \S\,\ref{mlre.sec}
this is partly due to an error in our
earlier analysis. The corrected
value is marginally
consistent with our new measurement ($\sigma=42^{+7}_{-7}$\,\kms\
from DEIMOS and $\sigma=33^{+3}_{-3}$\,\kms\ from KCWI), but we
cannot exclude other systematic effects. It may
be that the large weight of a Balmer line (H$\alpha$) in the analysis,
or the cross-talk corrections we had to apply, influenced the earlier
result.  The $M/L$ ratio
of \gal\ is $M_{\rm dyn}/L_I=26^{+7}_{-6}$\,M$_{\odot}/$L$_{\odot}$ within
the effective radius, and the galaxy is dominated by dark matter even in the center.
This does not necessarily mean that the galaxy has an ``overmassive'' halo;
as discussed in Fig.\ \ref{ml10.sec} and shown in Fig.\ \ref{wolf.fig} UDGs
are {\em expected} to have very high $M/L$ within the effective radius,
simply by virtue of being large. UDGs with ``normal'' $M/L(<R_{\rm e})$
ratios for
their luminosity, such as NGC\,1052-DF2, are the ones that deviate from
the expectations from the stellar mass -- halo mass relation.

We find that the velocity dispersion profile gradually increases with
radius. The profile cannot be fit with a standard NFW halo and an isotropic
velocity distribution: \gal\ either has a relatively flat density profile
(a core) or strong tangential anisotropy. The {Di Cintio} {et~al.} (2014a) model,
with a mass-dependent core, fits the data remarkably well. It reproduces
the observed velocity dispersion profile with isotropic orbits,
has a halo mass that is in very good agreement with the stellar mass -- halo
mass relation, and is in qualitative agreement with the positive $h_4$ 
parameter. Another way to phrase this result is that the kinematics
of \gal\ are similar to other galaxies in this stellar mass range, which
also show evidence for cores (see {Di Cintio} {et~al.} 2014b, 2014a, and references therein). Our results lend support to the model of
{Carleton} {et~al.} (2019), who show that cores may lead to ultra diffuse galaxy
formation in clusters as a result of tidal stripping. This model
is certainly consistent with the kinematics of \gal, but perhaps not
with its dynamically-cold local environment. It also remains to be
seen whether such tidal models
can explain the high globular cluster counts of \gal\ and other UDGs.

Irrespective of the detailed mass distribution it is clear that \gal\ has
a gravitationally-dominant dark matter halo, similar to many other
UDGs ({Beasley} {et~al.} 2016; {Toloba} {et~al.} 2018; {Mart{\'{\i}}n-Navarro} {et~al.} 2019), and
in apparent contrast to the UDGs
NGC\,1052-DF2 and NGC\,1052-DF4
(see, e.g., {van Dokkum} {et~al.} 2018b, 2019; {Martin} {et~al.} 2018; {Famaey}, {McGaugh}, \& {Milgrom} 2018; {Emsellem} {et~al.} 2019; {Danieli} {et~al.} 2019).
With a robust velocity dispersion measurement for NGC\,1052-DF2 from stellar
kinematics ({Danieli} {et~al.} 2019), the identification of a second galaxy in
the same class (NGC\,1052-DF4; {van Dokkum} {et~al.} 2019), and the results presented in this study,
there can be little doubt that
large, diffuse, spheroidal galaxies with stellar
masses of a few $\times 10^8$\,\msun\ have a remarkable range in their
kinematics and, hence, dark matter properties on kpc scales (see
Fig.\ \ref{wolf.fig}). This qualitatively addresses a point raised
by {Kroupa} (2012), who noted that the then-observed {\em low} scatter in
the SMHM relation is difficult to explain in the standard cosmological
model. Similar arguments have been made by {McGaugh} (2012)
on the basis of the Tully-Fisher relation.

The converse of this argument is
that a {\em high} scatter is difficult to explain in alternatives to dark matter,
such as Modified Newtonian Dynamics (MOND; {Milgrom} 1983) and Emergent Gravity
({Verlinde} 2017). The observed dispersion of \gal\ is higher than the MOND
prediction whereas the dispersions of NGC\,1052-DF2 and NGC\,1052-DF4 are lower.
Specifically, for a MOND acceleration scale of
$a_0=3.7\times 10^3$\,km$^2$\,s$^{-2}$\,kpc$^{-1}$ the predicted velocity dispersion
of \gal\ is $\sigma_{\rm M} \approx (0.05 GM_{\rm stars} a_0)^{1/4}
\approx 23$\,\kms, lower than the luminosity-weighted dispersion within
the effective radius of $\sigma_{\rm e}=34^{+3}_{-3}$\,\kms. 
The predicted
dispersions for NGC\,1052-DF2 and NGC\,1052-DF4 are
of the same order, whereas the
observed dispersions are $<10$\,\kms\ ({Danieli} {et~al.} 2019; {van Dokkum} {et~al.} 2019).
The ``external field effect'' ({Famaey} {et~al.} 2018) 
mitigates this tension, but it is difficult to explain dispersions as
low as $5-10$\,\kms\ for these
galaxies even when this effect is maximal (see {M{\"u}ller}, {Famaey}, \&  {Zhao} 2019).
A possible way to reconcile alternative dark matter models
with UDG kinematics is to invoke strong variations in the stellar
initial mass function (see, e.g., {Conroy} \& {van Dokkum} 2012; {Geha} {et~al.} 2013),
such that
\gal\ has a bottom-heavy mass function and the NGC\,1052 dwarfs are bottom-light.
There is no prior motivation for these specific variations, but they may be testable.

\begin{figure}[htbp]
  \begin{center}
  \includegraphics[width=0.95\linewidth]{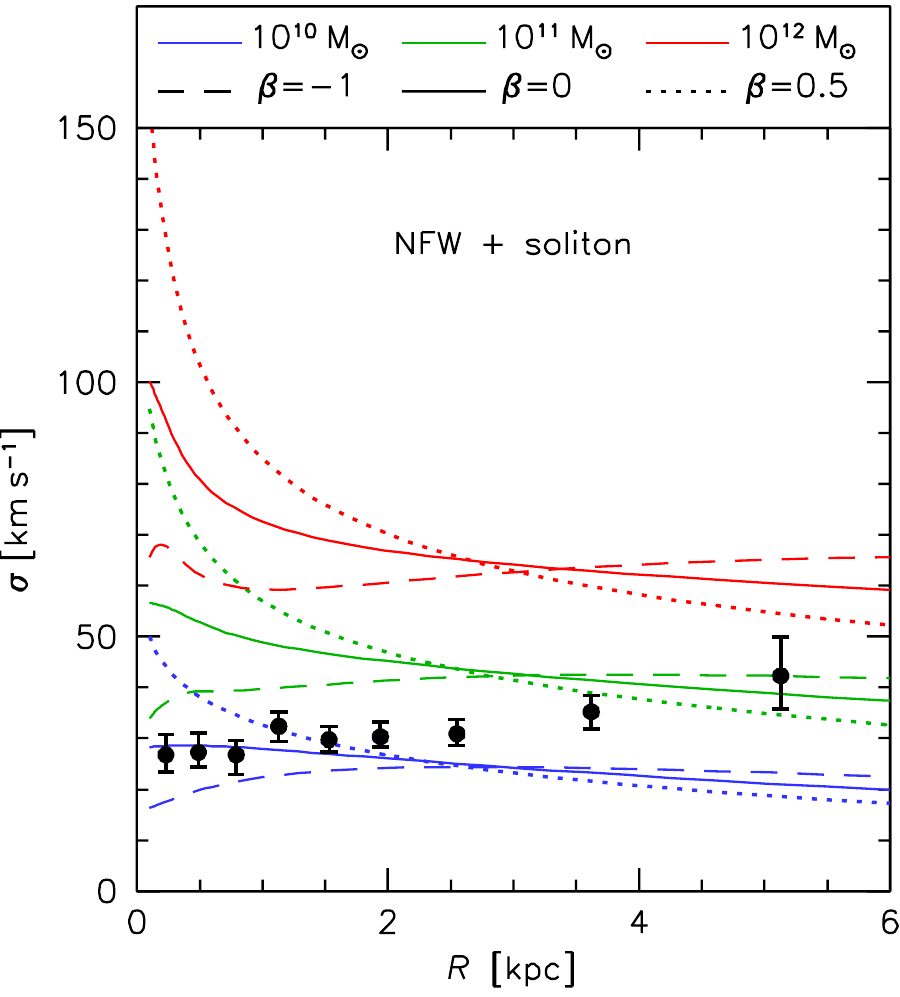}
  \end{center}
\vspace{-0.2cm}
    \caption{
Expected velocity dispersion profiles for ultra-light axion (``fuzzy'') dark
matter models. For sufficiently high halo masses these models predict a
characteristic bump in the profile at small ($\sim 500$\,pc) scales,
indicating the presence of a soliton core.
The mass of \gal\ is not quite high enough to determine whether NFW\,+\,soliton
models are preferred over standard NFW models. However, UDGs with higher central
dispersions may exist, and they could provide a direct test
of these predictions.
}
\label{fuzzy.fig}
\end{figure}

The fact that it is dark matter dominated on
all scales makes \gal, and other UDGs like it, important in the quest to
understand the distribution of dark matter on $\lesssim 1$~kpc scales. 
As noted above, there is a long history of using galaxies
with a low baryonic density
to constrain the density profiles of dark matter halos
(see, e.g., {Aaronson} 1983; {de Blok} {et~al.} 2001; {Kleyna} {et~al.} 2002; {Swaters} {et~al.} 2003, and many others).
As UDGs such as \gal\ are dark matter dominated on all scales,
as shown explicitly in Fig.\ \ref{fraction.fig},
they offer a ``pristine'' view of their dark matter even
on small spatial scales, which is
unusual for
galaxies with stellar masses of $M_{\rm stars}=10^8-10^9$\,\msun\ (and higher).
In particular, dwarf elliptical galaxies, ``classical'' low surface brightness disk
galaxies, and gas-rich dwarf galaxies all have
typical $M/L$ ratios in the range $5-10$
within the optical extent, and even lower values in the center
({Geha}, {Guhathakurta}, \& {van der  Marel} 2002; {Swaters} {et~al.} 2003; {Zaritsky} {et~al.} 2006; {Wolf} {et~al.} 2010).

A particularly interesting deviation in the dark matter profile occurs in
``fuzzy'' dark matter models, where the dark matter particle is an ultra-light
axion with a de Broglie wavelength of 100s of parsecs (e.g., {Marsh} \& {Silk} 2014).
At low halo masses the soliton core in these models is difficult to
distinguish, but at higher masses
($10^{11-12}$\,\msun) the predicted density profile can display a characteristic
bump on small scales. As pointed out by
{Hui} {et~al.} (2017), massive
UDGs may be able to constrain such models.
In Fig.\ \ref{fuzzy.fig} we show predicted velocity
dispersion profiles for
a galaxy with the surface brightness profile of \gal, using an NFW\,+\,soliton model
of the form proposed by {Marsh} \& {Pop} (2015).
We should not
expect a bump on 500\,pc scales in \gal, as the soliton core becomes
a distinct feature only for halo masses of $\gtrsim 10^{12}$\,\msun,
or velocity dispersions of $\sim 60$\,\kms. In that regime the effects
of the soliton can, with sufficiently accurate data, be distinguished from
those of anisotropy and variations in the halo mass (see also {Robles}, {Bullock}, \&  {Boylan-Kolchin} 2019).
In this context the recent announcement of a UDG with an apparent stellar
dispersion of $\sigma=56\pm 10$\,\kms\ ({Mart{\'{\i}}n-Navarro} {et~al.} 2019)
is exciting, as it suggests that
\gal\ does not  define the upper end of the halo mass range of UDGs.
We also note that, even though we cannot determine whether soliton models
provide a better fit than standard NFW models, we can place constraints on the
particle mass in the context of fuzzy dark matter models. These quantitative
constraints are given in  a companion paper (Wasserman et al.\ 2019).


\acknowledgements{
This paper is dedicated to Mariella Silvia, a young woman with a passion for
astronomy who has inspired us with her fortitude.
Support from {\em HST} grant HST-GO-14643
and NSF grants AST-1312376, AST-1616710, 
AST-1518294, and AST-1613582
is gratefully acknowledged. This work was partially supported by
a NASA Keck PI Data Award, administered by the NASA Exoplanet Science Institute.
The data presented herein were obtained at the
W.~M.~Keck Observatory, which is operated as a scientific partnership among the California Institute of Technology, the University of California and the National Aeronautics and Space Administration. The Observatory was made
possible by the generous financial support of the W.~M.~Keck Foundation. We are grateful to the staff of Keck
Observatory, and in particular Luca Rizzi, for their excellent support
and help. The authors wish to recognize and acknowledge the very significant cultural role and reverence that the summit of Maunakea has always had within the indigenous Hawaiian community.  We are most fortunate to have the opportunity to conduct observations from this mountain.
AJR is a Research Corporation for Science Advancement Cottrell Scholar.
}


\begin{appendix}
\section{A.\ Radial photometry}
In the sky subtraction process the surface brightness profile
of the galaxy is used to determine the absolute background level within the KCWI
field of view (see \S\,\ref{res.sec}). This profile is derived from the HST WFC3/UVIS $V_{606}$
image that was described in van Dokkum et al.\ (2017a). The central wavelength of the $V_{606}$
filter is $0.59\,\mu$m, slightly larger than the central wavelength of the KCWI spectrum ($0.51\,\mu$m),
and the sky subtraction could be in error if the galaxy has a strong color gradient.
Here we assess whether such a gradient exists, making use of the fact that \gal\ was
also observed in the $I_{814}$ filter (see van Dokkum et al.\ 2017a).

\begin{figure*}[htbp]
  \begin{center}
  \includegraphics[width=0.85\linewidth]{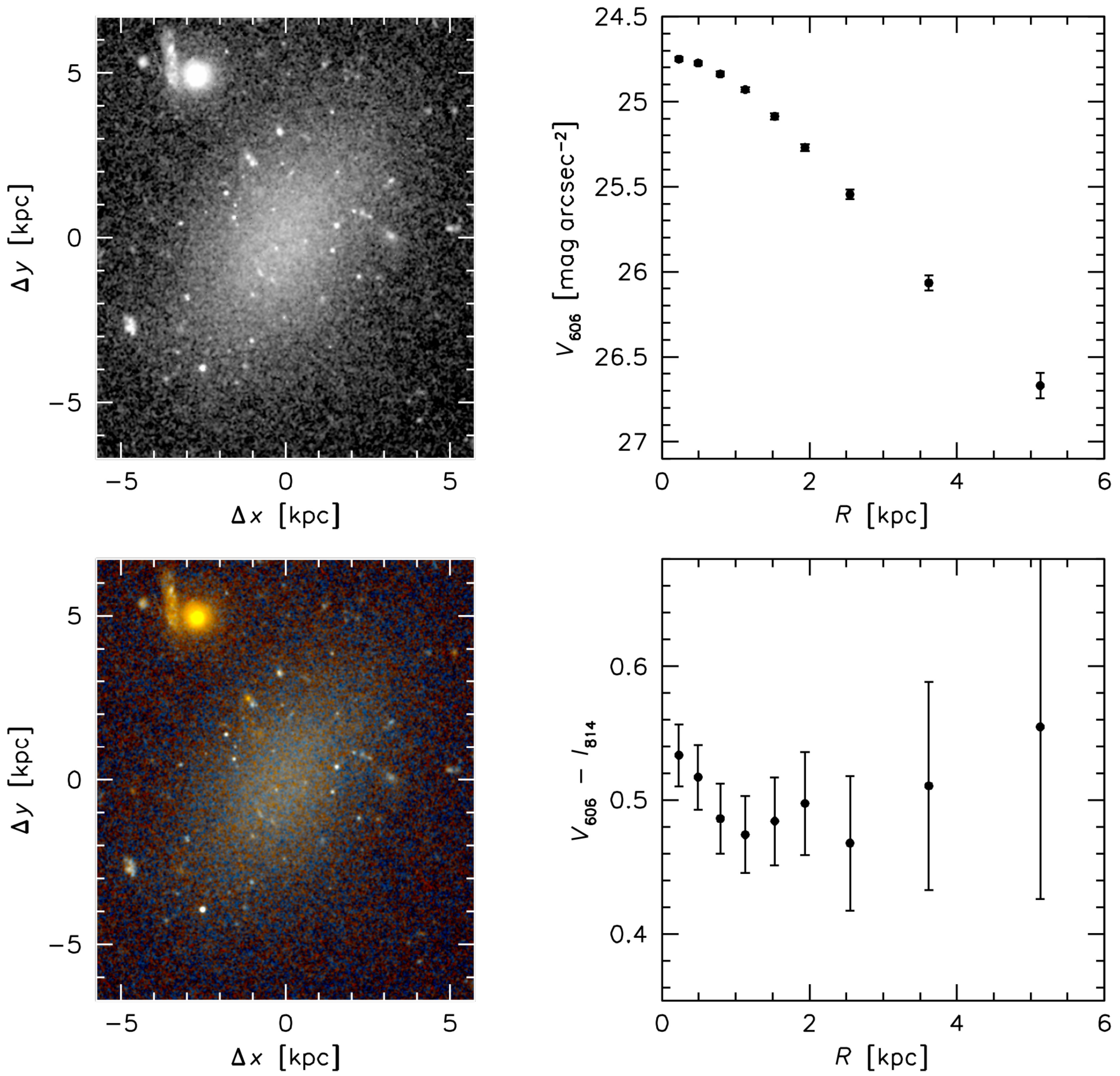}
  \end{center}
\vspace{-0.2cm}
    \caption{
Radial HST WFC3/UVIS photometry measured in the same apertures and at the same spatial
resolution as the KCWI kinematics. Top left: $V_{606}$ image, with North up and East
to the left. Top right: surface brightness profile in the elliptical apertures
of Fig.\ 6, after
degrading the $V_{606}$ image to the ground based resolution.
Bottom left: color image created from the $V_{606}$ and $I_{814}$ images.
Bottom right: color profile. The errors are dominated by the uncertainties in the background
level of the HST images.
}
\label{colorgrad.fig}
\end{figure*}

The $V_{606}$ image and surface brightness profile are shown in the top panels
of Fig.\ \ref{colorgrad.fig}. A color image
created from the $V_{606}$ and $I_{814}$ data is shown in the bottom
left panel.
To measure the color gradient
the $I_{814}$ image is smoothed to the ground-based seeing and resampled to the KCWI pixel scale,
mimicking the procedure that is used for the $V_{606}$ image. Next, $I_{814}$ fluxes are measured in the
same elliptical apertures (with the same masking of stars, globular clusters, and background
galaxies) as are used for the extraction of the KCWI spectra and the
measurements of the $V_{606}$ fluxes. Finally, colors are measured using the WFC3
zeropoints,\footnote{The WFC3/UVIS
$V_{606}$ and $I_{814}$ AB zeropoints are 26.103 and 25.139 respectively.}
with a $-0.01$ mag correction to account for Galactic reddening. Uncertainties are
dominated by the uncertainty in the background level in the HST images, and
determined from the variation in the mean flux in
empty apertures placed near \gal. The results are shown
in the bottom right panel of Fig.\ \ref{colorgrad.fig} and listed in Table \ref{colorgrad.table}.
There is no evidence for a gradient at $r>3\arcsec$, the part of the profile that is used
to scale the profile derived from the KCWI data.

\begin{deluxetable*}{cccc}
\tablecaption{Radial Photometry
\label{colorgrad.table}}
\tablehead{\colhead{$R$} & \colhead{$R$}  & \colhead{$V_{606}$} & \colhead{$V_{606}-I_{814}$} \\
\colhead{[arcsec]}& \colhead{[kpc]} & \colhead{[mag\,arcsec$^{-2}$]} & \colhead{[mag]}}
\startdata
$0\farcs 5$ & 0.23 & $24.75 \pm 0.01$ & $0.53\pm 0.02$ \\
$1\farcs 0$ & 0.49 & $24.77 \pm 0.01$ & $0.52 \pm 0.02$ \\
$1\farcs 6$  & 0.79 & $24.84 \pm 0.01$ & $0.49 \pm 0.03$ \\
$2\farcs 3$  & 1.13 & $24.93 \pm 0.02$ & $0.47 \pm 0.03$\\
$3\farcs 2$  & 1.53 & $25.09 \pm 0.02$ & $0.48\pm 0.03$ \\
$4\farcs 0$  & 1.94 & $25.27 \pm 0.02$ & $0.50 \pm 0.04$ \\
$5\farcs 3$  & 2.55 & $25.54 \pm 0.03$ & $0.47 \pm 0.05$ \\
$7\farcs 4$ & 3.62 &  $26.07 \pm 0.04$ & $0.51 \pm 0.08$ \\
$10\farcs 6$ & 5.13 & $26.67 \pm 0.08$ & $0.55 \pm 0.13$ 
\enddata
\end{deluxetable*}

\end{appendix}

\end{document}